\documentclass[structabstract]{aa}

\usepackage{natbib}
\bibpunct{(}{)}{;}{a}{}{,} 

\usepackage{graphicx}

{}

\usepackage{txfonts}

\newcommand{\eqref}[1]{Eq.\,(\ref{#1})}

\newcommand{\figref}[1]{Fig.\,\ref{#1}}
\newcommand{\tabref}[1]{Tab.\,\ref{#1}}
\newcommand{\secref}[1]{Sect.\,\ref{#1}}
\newcommand{\Alfven}{Alfv{\'e}n}

\newcommand{\pns}{proto-neutron star}

\newcommand{\Pst}{P_{\star}}
\newcommand{\est}{e_{\star}}

\newcommand{\ms}{\mathrm{ms}}
\newcommand{\km}{\mathrm{km}}

\newcommand{\cms}{\mathrm{cm\,s}^{-1}}

\newcommand{\gccm}{\mathrm{g\,cm}^{-3}}

\newcommand{\msun}{M_{\sun}}

\newcommand{\Msol}{M_{\sun}}

\newcommand{\Alnum}{\mathsf{A}}

\newcommand{\neutrino}[1]{{#1}_{\nu}}

\newcommand{\modelname}[1]{\texttt{#1}}
\newcommand{\subpanel}[1]{\textit{#1}}

\begin{document}

\title{
  Magnetic field amplification in collapsing, \\ non-rotating stellar
  cores
}
{\author{
    M.~Obergaulinger\inst{\ref{MPA},\ref{HUJI}}
    \and 
    H.-Th.~Janka\inst{\ref{MPA}}
  }
  \institute{
    Max-Planck-Institut f{\"u}r Astrophysik, Karl-Schwarzschild-Str.~1,
    D-85748 Garching, Germany\label{MPA}
   \and
   Racah Institute of Physics, The Hebrew University, Jerusalem 91904,
   Israel\label{HUJI}
  }
  \date{Received xx month xxxx / Accepted xx month xxxx}
}

{\abstract
  {
    The influence of magnetic fields on stellar core collapse and
    explosion is not well explored. It depends on the possibility to
    amplify the initial pre-collapse fields.  In the absence of
    rotation this can happen by compression, convection, the standing
    accretion shock instability, and the accumulation and growth of
    \Alfven~waves in the accretion flows.
  }
  {
    We investigate such amplification mechanisms of the magnetic
    field during the collapse and post-bounce evolution of the core
    of a non-rotating 15$\,\msun$ star with varied initial field
    strengths, taking into account the microphysical equation of state
    and neutrino physics that play a crucial role in supernova cores.
  }
  {
    We perform simulations of ideal magnetohydrodynamics with neutrino
    transport in axisymmetry. The transport of electron neutrinos and
    antineutrinos is treated with a new scheme that solves the
    energy-dependent set of radiation energy and momentum equations in
    two dimensions by using an analytic closure relation.
  }
  {
    The magnetic field undergoes kinematic amplification by turbulent
    flows. We also find indications for amplification by interacting
    waves travelling upwards and downwards inside accretion
    streams. The fields can reach up to equipartition with the
    velocity field. Very high magnetic field strengths require very
    strong pre-collapse fields and are able to shape the post-bounce
    flow, leading to a pattern dominated by low-order multipoles. Such
    models are closest to a successful explosion.
  }
  {
    Magnetic fields can build up to interesting strengths even in
    non-rotating collapsing stellar cores. Starting with fields in the
    pre-collapse core as predicted by present stellar evolution models
    ($10^{9}$--$10^{10}$\,G), typical neutron star fields of
    $10^{12}$--$10^{13}$\,G emerge, whereas progenitor fields of some
    $10^{11}$--$10^{12}$\,G lead to fields of magnetar strength of
    about $10^{14}$--$10^{15}$\,G. Only in the latter case the
    magnetic fields have significant dynamical effects on the flows in
    the supernova core and may have an influence on the explosion
    mechanism.  However, in none of our 2D simulations, we find an
    explosion until 500 milliseconds post-bounce.
  }
}

\maketitle


\section{Introduction}
\label{Sec:Intro}

Most scenarios for the explosion mechanism of core-collapse supernovae
(SNe) involve a combination of energy deposition in the matter
surrounding the nascent \pns~(PNS) and multi-dimensional hydrodynamic
flows.  Examples for means of energy transfer to the SN ejecta are the
prompt bounce shock, neutrinos, magnetic fields and acoustic waves.
In the neutrino-heating mechanism neutrinos tap the gravitational
potential energy released during collapse and deposit a part of it
behind the stalled shock. This process is enhanced and thus supported
by non-radial fluid flows triggered by hydrodynamic instabilities like
convection and the standing accretion shock instability
\citep[SASI;][]{Blondin_Mezzacappa_DeMarino__2003__apj__Stability_of_Standing_Accretion_Shocks_with_an_Eye_toward_Core-Collapse_Supernovae,Foglizzo__2001__aap__Entropic-acoustic_instability_of_shocked_Bondi_accretionI._What_does_perturbed_Bondi_accretion_sound_like?,Foglizzo__2002__aap__Non-radial_instabilities_of_isothermal_Bondi_accretion_with_a_shock:Vortical-acoustic_cycle_vs.post-shock_acceleration}

Though it is safe to assume that the exploding star will possess a
magnetic field of some (unknown) strength and topology, its possible
influence on the explosion is less clear.  The main reason for the
small number of conclusive investigations into this topic is that
important effects are expected to occur only when the magnetic field
is roughly in (energetic) equipartition with the gas flow, a condition
corresponding to extremely strong fields similar to those observed in
magnetars.  This makes full MHD simulations including a treatment of
the important neutrino physics in the SN core indispensable.

Stellar evolution calculations, predicting only weak pre-collapse
magnetic fields, render the prospects for magnetically affected
explosions very much dependent on the amount of field amplification
happening during and after collapse.  Rapid rotation may amplify a
weak seed field quite generically to dynamically relevant values,
e.g., by winding up a poloidal field (a process linear in time) or,
exponentially in time, by the magneto-rotational instability
\citep[MRI,][]{Akiyama_etal__2003__ApJ__MRI_SN}.  Magneto-rotational
explosions, theoretically discussed by
\cite{Meier_etal__1976__ApJ__MHD_SN}, have been studied in various
approximations, e.g., by
\cite{Bisnovatyi-Kogan_Popov_Samokhin__1976__APSS__MHD_SN,Symbalisty__1984__ApJ_MHD_SN,Akiyama_etal__2003__ApJ__MRI_SN,Kotake_etal__2004__Apj__SN-magrot-neutrino-emission,Obergaulinger_Aloy_Mueller__2006__AA__MR_collapse,Cerda-Duran_et_al__2007__AA__passive-MHD-collapse,
  Obergaulinger_etal__2009__AA__Semi-global_MRI_CCSN}; recent
simulations employing detailed microphysics have been performed by
\cite{Burrows_etal__2007__ApJ__MHD-SN,Scheidegger_etal__2008__aap__GW_from_3d_MHD_SN}.

However, according to stellar-evolution models, the majority of
progenitors is expected to rotate slowly.  Although many stars on the
upper main sequence show a surface rotation period close to the
critical value for mass shedding, they will most likely lose most of
their angular momentum during their subsequent evolution, e.g., by
strong stellar winds or magnetic braking
\citep{Heger_et_al__2005__apj__Presupernova_Evolution_of_Differentially_Rotating_Massive_Stars_Including_Magnetic_Fields,Meynet_etal__2011__aap__Massive_star_models_with_magnetic_braking}.
Non-rotating cores, albeit lacking the above-mentioned efficient
channels for amplification, may experience field growth by several
effects:

\begin{enumerate}
\item After bounce, convection develops in the PNS and the surrounding
  hot-bubble region.  Breaking down into three-dimensional turbulence,
  convection may provide the $\alpha$ effect responsible for a
  small-scale dynamo amplifying the field on length scales comparable
  to the scale of turbulent forcing, i.e., the size of convective
  eddies \citep[][]{Thompson_Duncan__1993__ApJ__NS-dynamo}.  However,
  the generation of a large-scale field, e.g., on the scale of the
  PNS, probably requires a non-vanishing kinetic helicity of the
  turbulent flow, which can most naturally be accounted for by
  (differential) rotation \citep[see,
  e.g.,][]{Brandenburg_Subramanian__2005__AN__Strong_mean_field_dynamos_require_supercritical_helicity_fluxes}.
\item
  \citet{Endeve_et_al__2010__apj__Generation_of_Magnetic_Fields_By_the_SASI}
  have demonstrated that the standing-accretion-shock instability
  \citep[SASI][]{Blondin_Mezzacappa_DeMarino__2003__apj__Stability_of_Standing_Accretion_Shocks_with_an_Eye_toward_Core-Collapse_Supernovae},
  excited by a feedback cycle of acoustic and advective perturbations
  travelling between the shock wave and the deceleration region above
  the PNS, has the potential to amplify the magnetic field by up to
  four orders of magnitude.  Dynamically relevant field strengths (in
  which case the field reaches at least 10\% of the equipartition
  strength) can be reached only when the pre-collapse field in the
  stellar core is sufficiently strong.
\item Non-radial fluid motions triggered by these instabilities can
  excite perturbations of the magnetic field propagating as
  \Alfven~waves along the field lines.  The outward propagation of
  \Alfven~waves excited close to the PNS has to compete with the
  accretion of gas towards the centre.  Assuming that the accretion
  flow decelerates in this region continuously,
  \cite{Guilet_et_al__2010__ArXive-prints__Dynamics_of_an_Alfven_surface_in_CCSNe}
  argue that there must be an \emph{\Alfven~point} at which the
  \Alfven~speed equals the accretion velocity, and the propagation of
  the wave (measured in the lab frame) comes to a rest.  They show
  that \Alfven~waves are amplified exponentially at such a stagnation
  point.  At the conditions of a supernova core, the amplification
  should be most efficient for a magnetic field strength of a few
  $10^{13}$ G and could yield final fields of the order of $10^{15}$
  G.  Dissipation of the wave energy can increase the entropy of the
  gas, thus modifying the dynamics in the accretion region.  For even
  stronger fields, the \Alfven~point can be close to the shock wave.
  In this case,
  \cite{Suzuki_Sumiyoshi_Yamada__2008__ApJ__Alfven_driven_SN} find
  that the explosion can be driven solely by the energy deposited by
  the dissipation of \Alfven~waves.  This process transmitting energy
  from the (convectively active) PNS to the much less dense
  surrounding medium bears a strong similarity to the proposed
  mechanism for heating the solar corona by \Alfven~waves emerging
  from the solar surface.
\end{enumerate}

Previous simulations of magnetohydrodynamic stellar core collapse have
used a wide variety of methods to treat the effects of neutrinos: in
the simplest models, they were either ignored
\citep[e.g.,][]{Obergaulinger_Aloy_Mueller__2006__AA__MR_collapse,Mikami_et_al__2008__apj__3d_MHD_SN}
or treated by simple parameterizations or local source terms
\citep[e.g.,][]{Cerda-Duran__2008__AA__GRMHD-code}; more complex
approaches include trapping/leakage schemes
\citep[e.g.,][]{Kotake_etal__2004__Apj__SN-magrot-neutrino-emission},
multi-dimensional, energy-dependent flux-limited diffusion
\citep[][]{Dessart_et_al__2006__apj__Multi-d_RHD_Simulations_of_PNS_Convection}
or the isotropic-diffusion source approximation which discriminates
between trapped and free-streaming components
\citep[][]{Liebendorfer_et_al__2009__apj__The_Isotropic_Diffusion_Source_Approximation_for_Supernova_Neutrino_Transport}.
In our approach to radiation-magnetohydrodynamics (RMHD), we employ a
new multi-dimensional and energy-dependent scheme for the neutrino
transport in supernova cores, namely a two-moment solver for the
neutrino energy (lepton number) and momentum equations with an
analytic closure relation
\citep[][]{Cernohorsky_van_Weert__1992__ApJ__Rel_2_moment_nu,Pons_Ibanez_Miralles__2000__MNRAS__hyperbol_radtrans,Audit_et_al__2002__astro-ph__hyp_RHD_closure}. Presently
it includes only electron neutrinos and antineutrinos and a relatively
simple treatment only of the most relevant interaction rates of these
neutrinos with neutrons, protons, and nuclei.  Nevertheless, in the
context of the questions focussed on in this paper, the scheme
provides a reasonably good representation of the neutrino effects that
play a role during core collapse, bounce, shock propagation and in the
accretion layer behind the stalled supernova shock. Moreover, it is a
computationally efficient treatment of the neutrino transport, and
two-dimensional simulations up to several hundred milliseconds after
bounce are easily possible.

It is the goal of this work to investigate the relevance of field
amplification mechanisms like those in points 1.-3.~concerning their
importance for the evolution and potential revival of the stalled
shock by core-collapse simulations including a reasonably good
treatment of the relevant microphysics.  In particular, we will study
the field growth connected to convective and SASI activity in the
post-shock layer and the role of energy transport and dissipation by
\Alfven~waves.  To this end, we perform MHD simulations of collapse
and post-bounce evolution of the core of a star of 15 solar masses
\citep[][]{Woosley_Heger_Weaver__2002__ReviewsofModernPhysics__The_evolution_and_explosion_of_massive_stars}.
Varying the strength of the initial field, assumed to be purely
poloidal, we can determine different sites and mechanisms of field
amplification and identify the back-reaction of the field onto the
flow.

This article is organised as follows: \secref{Sek:Phys-Num} describes
our physical model and the numerical methods; \secref{Sek:InCond}
introduces the initial conditions; \secref{Sek:Res} presents the
results of our simulations; \secref{Sek:SumCon} gives a summary of the
study and draws some conclusions; appendix \ref{Sek:App-NT} gives a
brief overview of our treatment of the neutrino transport.

\section{Physical model and numerical methods}
\label{Sek:Phys-Num}

\subsection{Magnetohydrodynamics}
\label{sSek:MHD}

\begin{table}
  \centering
  \caption{
    Variables and symbols used in the MHD equations and the neutrino
    transport formulation presented in this article.
  }
  \begin{tabular}{c|l}
    Symbol & Variable
    \\
    \hline
    $\rho$ & gas density
    \\
    $Y_{\mathrm{e}}$ & electron fraction
    \\
    $\vec v$ & gas velocity
    \\
    $\vec b$ & magnetic field
    \\
    $\varepsilon$ & internal energy density
    \\
    $\est$ & total energy density
    \\
    $P$ & gas pressure
    \\
    $\Pst$ & total pressure
    \\
    $\Phi$ & gravitational potential
    \\
    $\neutrino{S}^0$ & neutrino-matter energy exchange
    \\
    $S_{\nu; \mathrm{n}}^0$ & neutrino-matter lepton-number exchange
    \\
    $\neutrino{S}^i$ & neutrino-matter momentum exchange
    \\
    $\neutrino{E}$ & neutrino energy density
    \\
    $\neutrino{N}$ & neutrino number density
    \\
    $\neutrino{F}^{i}$ & neutrino energy flux
    \\
    $\neutrino{V}^{i}$ & neutrino number flux
    \\
    $\neutrino{p}$ & scalar Eddington factor
    \\
    $\neutrino{P}^{ij}$ & neutrino energy pressure tensor
    \\
    $\neutrino{R}^{ij}$ & neutrino number pressure tensor
    \\
    $\chi^0$ & $0^{\mathrm{th}}$-moment opacity
    \\
    $\chi^1$ & $1^{\mathrm{st}}$-moment opacity
  \end{tabular}
  \label{Tab:MHD-Sym}
\end{table}

We assume that the evolution of the gas and the magnetic field is
described by the equations of Newtonian ideal magnetohydrodynamics
(MHD),
\begin{eqnarray}
  \label{Gl:MHD-rho}
  \partial_{t} \rho + \vec \nabla \cdot (\rho \vec v) 
  & = & 0,
  \\
  \label{Gl:MHD-y_e}
  \partial_{t} (\rho Y_{\mathrm{e}}) + \vec \nabla \cdot (\rho Y_{\mathrm{e}} \vec v )
  & = & S_{\nu; \mathrm{n}}^0,
  \\
  \label{Gl:MHD-mom}
  \partial_{t} (\rho v^i )
  + \nabla_j \left( \Pst \delta^{ij} + \rho v^i v^j - b^i b^j
    \right )
    & = &
    \rho \nabla^i \Phi
    + \neutrino{S}^i,
    \\
    \label{Gl:MHD-erg}
    \partial_{t} \est 
    + \vec \nabla \cdot 
    \left(
      (\est + \Pst) \vec v - (\vec v \cdot \vec b ) \vec b
    \right)
    & = &
    \rho \vec v \cdot \vec \nabla \Phi
    + \neutrino{S}^0,
    \\
    \label{Gl:MHD-ind}
    \partial_{t} \vec b 
    - 
    \vec \nabla \times
    \left(
      \vec v \times \vec b
    \right)
    & = &
    0,
\end{eqnarray}
describing the conservation of mass, electron fraction, gas momentum,
total energy of the matter, and magnetic flux, respectively.  In
addition to these evolutionary equations, the magnetic field has to
fulfil the divergence constraint,
\begin{equation}
  \label{Gl:MHD-divb}
  \vec \nabla \cdot \vec b = 0.
\end{equation}
The symbols used in this system have standard meaning; they are listed
in \tabref{Tab:MHD-Sym}.  The total energy density and pressure are
defined as the sum of the contributions of the gas and the magnetic
field, $\est = \varepsilon + 1/2 \, \rho \vec v^2 + 1/2 \, \vec b^2$
and $\Pst = P + 1/2 \, \vec b^2$, respectively.

The MHD part of the system is closed by the equation of state of
\citet{Shen_et_al__1998__PThPh__Rel_EOS}, relating gas density,
internal-energy density, and electron fraction to temperature, $T$,
gas pressure $P$, entropy, $S$, the composition of four species
(neutrons, protons, $\alpha$ particles, and a representative heavy
nucleus), and the chemical potentials of neutrons, protons, and
electrons.

To include some key features of general relativity into our Newtonian
model, we use the modified gravitational potential of
\citet{Marek_etal__2006__AA__TOV-potential}.  Following the lines
proposed by these authors, we include the gas and neutrino energy
density and pressure as well as an approximate Lorentz factor of the
flow in the source term of the Poisson equation for $\Phi$.  Overall,
this leads to a somewhat deeper potential well when the core
approaches nuclear matter density and serves as a good approximation
to the stronger relativistic gravity of a PNS.

Our code is based on an implementation of the MHD equations on a fixed
Eulerian grid in the constrained-transport discretisation
\citep{Evans_Hawley__1998__ApJ__CTM}.  It uses \emph{high-resolution
  shock-capturing (HRSC)} methods with approximate Riemann solvers
(Lax-Friedrichs, HLL, or HLLD
\citep[][]{Miyoshi_Kusano__2005__JCoP__HLLD_for_ideal_MHD}) in the
multi-stage framework \citep{Titarev_Toro__2005__IJNMF__MUSTA}.  We
reconstruct the conserved variables using high-order
\emph{monotonicity-preserving (MP)} methods
\citep[][]{Suresh_Huynh__1997__JCP__MP-schemes}.  Time stepping is
done by explicit Runge-Kutta schemes of $2^{\mathrm{nd}}$,
$3^{\mathrm{rd}}$, or $4^{\mathrm{th}}$ order.

\subsection{Neutrino transport and neutrino-matter interactions}
\label{sSek:RT}

We treat the transport of  neutrinos by solving the energy-dependent,
multi-dimensional moment equations for neutrino energy,
$\neutrino{E}$, and neutrino momentum, $\neutrino{F}^i$,
\begin{eqnarray}
  \label{Gl:RT-0mom}
  \partial_{t} \neutrino{E} 
  + \vec \nabla \cdot ( \neutrino{E} \vec v + \neutrino{\vec F})
  - \omega \nabla_j v_k \partial_{\omega} \neutrino{P}^{jk}
  & = &
  \neutrino{S}^0
  ,
  \\
  \label{Gl:RT-1mom}
  \partial_{t} \neutrino{F}^i 
  + \nabla_j ( \neutrino{F}^i v^j + \neutrino{P}^{ij} )
  + \neutrino{F}^i \nabla_j v^j
  & = &
  \neutrino{S}^{1; i}
  .
\end{eqnarray}
Presently only electron neutrinos, $\nu_{\mathrm{e}}$, and
antineutrinos, $\bar{\nu}_{\mathrm{e}}$ are considered.  The source
terms, $\neutrino{S}^{0,1}$, include the interaction of neutrinos with
the gas by emission, absorption, and scattering reactions; we use a
reduced set of processes, viz.
\begin{enumerate}
\item emission and absorption of electron neutrinos by neutrons,
\item emission and absorption of electron anti-neutrinos by protons,
\item elastic scattering of all neutrinos off nucleons,
\item emission and absorption of electron neutrinos by heavy nuclei,
\item coherent elastic scattering of all neutrinos off heavy nuclei.
\end{enumerate}
Furthermore, fluid-velocity dependent effects such as the $P \,
\mathrm{d} V$ work associated with diverging flows appear in the
equations as well.  To close the system of moment equations, we have
to specify the tensor of the second moment, $\neutrino{P}^{ij}$, for
which we use a simple analytic closure relation from
\cite{Minerbo__1978__jqsrt__Maximum_entropy_Eddington_factors}.  Using
a tensorial generalisation of a one-dimensional \emph{Eddington
  factor}, our method is generically multi-dimensional.  A more
detailed discussion can be found in Appendix \ref{Sek:App-NT}.

We note that our set of equations is a generalisation of the usual
diffusion ansatz, in which the system of moment equations is truncated
at the level of the energy equation and closed by expressing the flux
in terms of the gradient of the energy density, mostly connecting
diffusion and free-streaming limits by a flux limiter.  Retaining the
first two moments, it leads to a hyperbolic system, which can be
solved by common methods such as high-resolution shock-capturing
methods.  More details on this can be found in Appendix
\ref{Sek:App-NT}.

\section{Models and initial conditions}
\label{Sek:InCond}

\begin{table}
  \centering
  \caption{
    Average magnetic field strengths and normalised magnetic energies for our
    computed models with magnetic field (excluding model
    \modelname{s15-B0}).  See text for details.
  }
  \begin{tabular}{l|cccc}
    \hline
    model 
    & \modelname{s15-B10}
    & \modelname{s15-B11}
    & \modelname{s15-B11.5}
    & \modelname{s15-B12}
    \\ 

    \hline
    \hline
    $b_0$
    & $10^{10}$
    & $10^{11}$
    & $3.16 \times 10^{11}$
    & $10^{12}$
    \\ 
    \hline
    $b_{14}$ [G] 
    & $1.2 \times 10^{13}$
    & $1.4 \times 10^{14}$
    & $4.2 \times 10^{14}$
    & $1.4 \times 10^{15}$
    \\ 
    $\beta^{\mathrm{i}}_{14}$
    & $5.9 \times 10^{-9}$
    & $7.8 \times 10^{-7}$
    & $7.5 \times 10^{-6}$
    & $5.9 \times 10^{-5}$
    \\ 
    $\beta^{\mathrm{k}}_{14}$
    & $0.20$
    & $24$
    & $82$
    & $700$
    \\ 

    \hline
    $b_{\mathrm{cnv}}$ [G] 
    & $9.1 \times 10^{12}$
    & $1.2 \times 10^{14}$
    & $2.1 \times 10^{14}$
    & $4.7 \times 10^{14}$
    \\ 
    $\beta^{\mathrm{i}}_{\mathrm{cnv}}$
    & $1.1 \times 10^{-7}$
    & $2.9 \times 10^{-5}$
    & $9.8 \times 10^{-5}$
    & $4.8 \times 10^{-4}$
    \\ 
    $\beta^{\mathrm{k}}_{\mathrm{cnv}}$
    & 0.00041
    & 0.055
    & 0.16
    & 1.6
    \\ 

    \hline
    $b_{\mathrm{stb}}$ [G] 
    & $3.8 \times 10^{12}$
    & $3.0 \times 10^{13}$
    & $6.3 \times 10^{13}$
    & $1.4 \times 10^{14}$
    \\ 
    $\beta^{\mathrm{i}}_{\mathrm{stb}}$
    & $1.1 \times 10^{-6}$
    & $7.8 \times 10^{-5}$
    & $3.3 \times 10^{-4}$
    & $1.4 \times 10^{-3}$
    \\ 
    $\beta^{\mathrm{k}}_{\mathrm{stb}}$
    & 0.0020
    & 0.16
    & 0.80
    & 3.7
    \\ 

    \hline
    $b_{\mathrm{gain}}$ [G] 
    & $3.6 \times 10^{11}$
    & $6.0 \times 10^{12}$
    & $1.0 \times 10^{13}$
    & $1.2 \times 10^{13}$
    \\ 
    $\beta^{\mathrm{i}}_{\mathrm{gain}}$
    & $2.2 \times 10^{-6}$
    & $5.7 \times 10^{-4}$
    & $2.7 \times 10^{-3}$
    & $1.0 \times 10^{-2}$
    \\ 
    $\beta^{\mathrm{k}}_{\mathrm{gain}}$
    & 0.000089
    & 0.016
    & 0.071
    & 0.26
    \\

    \hline
    \hline

  \end{tabular}

  \label{Tab:models}
\end{table}

\begin{figure}
  \centering
  \resizebox{\hsize}{!}{\includegraphics{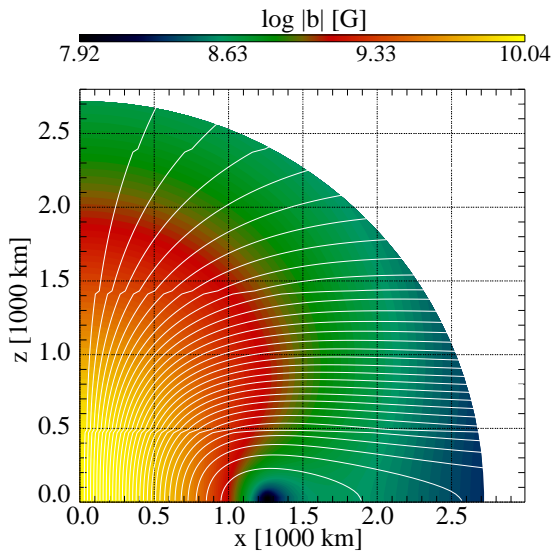}}
  \caption{
    Initial (pre-collapse) field configuration of model
    \modelname{s15-B10}: field strength (colour scale) and field
    lines.  For simplicity, only one quadrant is shown.  }
  \label{Fig:B-init}
\end{figure}

Interested in the contribution of magnetic fields to the dynamics of
supernovae, we simulate the evolution of a star of 15.0 solar masses
with solar metalicity
\citep[][]{Woosley_Heger_Weaver__2002__ReviewsofModernPhysics__The_evolution_and_explosion_of_massive_stars}.
We map the pre-collapse model to a grid of $n_r = 320$ radial zones
distributed logarithmically over a radial domain of $[r_{\mathrm{min}}
=0; r_{\mathrm{max}} \approx 2800 ~ \mathrm{km}]$ and $n_{\theta} =
64$ lateral zones distributed uniformly between the north and south
pole.  The resolution at the grid centre is $(\delta r)_{\mathrm{ctr}}
= 400~\mathrm{m}$.  Ten energy bins between $\epsilon_{\mathrm{min}} =
0.1~\mathrm{MeV}$ and $\epsilon_{\mathrm{max}} = 280~\mathrm{MeV}$ are
used for the neutrinos.

The topology of the magnetic field at the onset of collapse is highly
uncertain.  On the main sequence, field amplification by, e.g.,
gradual contraction of the star or convection competes with the loss
of magnetic energy in stellar winds and in work the magnetic field
does by exerting torques on the stellar matter.  In the absence of
rotation, stars lack an important ingredient of most large-scale
dynamos. In such a case, they may be dominated by a small-scale
turbulent field rather than a large-scale field, e.g., a lower-order
multipole.  Nevertheless, we assume a simple initial field, viz.~a
modified dipole (\figref{Fig:B-init}).  While this is probably not a
very good approximation to real stellar cores, it represents an
optimal configuration for the \Alfven-wave amplification mechanism we
are interested in because of the large coherent radial component
favouring the radial propagation of \Alfven~waves.  Furthermore, we
observe that the field is replaced by a more complex small-scale field
in the regions of the core affected by hydrodynamic instabilities such
as convection and the SASI.  Hence, we deem the influence of such an
artificial choice for the initial field on our results only a minor
one.

The field is the same as the one used by
\cite{Suwa_etal__2007__pasj__Magnetorotational_Collapse_of_PopIII_Stars},
defined by a vector potential of the form
\begin{equation}
  \label{Gl:Init-A}
  A^{\phi} = \frac{b_0}{2} \frac{r_0^3}{r^3 + r_0^3} r \sin \theta.
\end{equation}
We set the normalisation radius $r_0$ denoting the location of the
dipole to $r_0 = 1000~\mathrm{km}$ and vary the parameter $b_0$
setting the field at the centre of the core between $10^{10}$ and
$10^{12}$ G, i.e., in a range somewhat above that found by
\citep{Heger_et_al__2005__apj__Presupernova_Evolution_of_Differentially_Rotating_Massive_Stars_Including_Magnetic_Fields}
for rotating progenitors.  See \figref{Fig:B-init} for a visualisation
of the initial field of model \modelname{s15-B10}.

\tabref{Tab:models} list important parameters of our models: apart
from the initial field strength, $b_0$, we present time averages (over
$t \in [280 ~ \ms, 300 \ms]$) of the mean field strength in the PNS at
densities above $10^{14} ~ \gccm$, in the convection zone inside the
PNS ($b_{\mathrm{cnv}}$), in the stable layer surrounding the PNS
($b_{\mathrm{stb}}$), and in the gain region where neutrinos deposit
energy behind the SN shock ($b_{\mathrm{gain}}$).  Moreover,
time-averaged (over $t \in [280 \, \ms,300 \, \ms]$) ratios of
magnetic energy to internal energy and to kinetic energy in the same
regions, $\beta^{\mathrm{i}}$ and $\beta^{\mathrm{k}}$, respectively,
are listed.  We will define these regions more quantitatively in
\secref{sSek:Res-B0}.

\section{Results}
\label{Sek:Res}

\subsection{A non-magnetised reference model}
\label{sSek:Res-B0}

\begin{figure*}
  \centering
  \includegraphics[width=17cm]{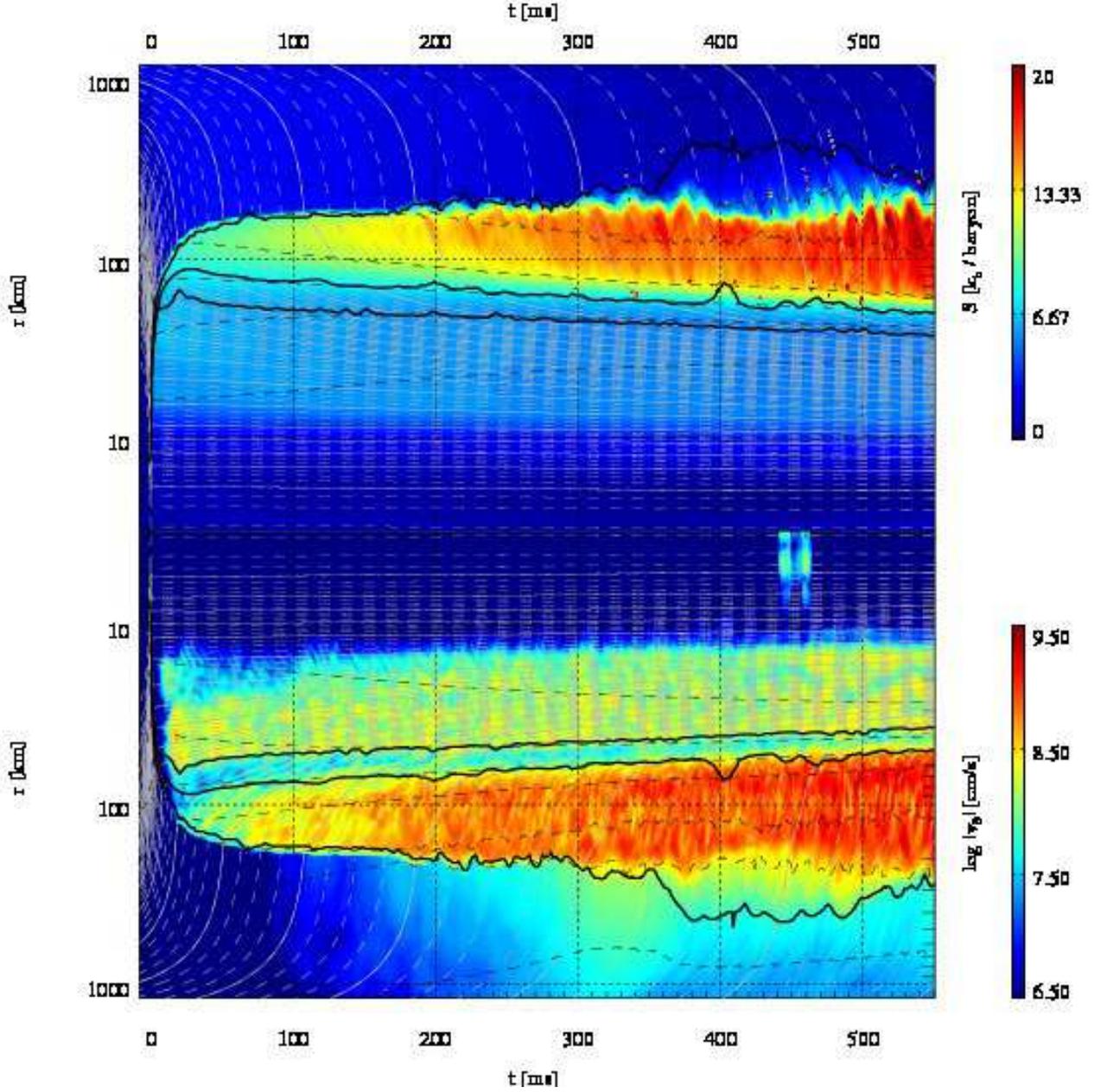}
  \caption{
    Summary of the evolution of model \modelname{s15-B0}.  The upper
    and lower panels show angular averages of the entropy per baryon
    and the of root-mean-square of the lateral velocity, respectively
    (colour coding) as a function of radius and post-bounce time.  The
    grey lines indicate the positions of different mass shells; the
    intervals between solid and dashed lines are $0.1$ and $0.025$
    solar masses, respectively.  The thick black lines show (from
    small to large radii) the outer boundaries of the PNS convection
    zone and of the stable layer beneath the SASI region and the
    outermost position of the shock wave.  The thin dashed black lines
    depict iso-density surfaces of $10^n ~ \gccm, n = 14,13,...$.  }
  \label{Fig:s15.0--Om0--B0--mshells}
\end{figure*}

The model with $b_0 = 0$, model \emph{\modelname{s15-B0}}, serves as a
reference case to which we can compare the magnetised models;
furthermore, it allows us to compare our results with previous results
for similar cores.  The evolution of the mean entropy profile and of
the root-mean-squared lateral velocity of the model as function of
time is shown in \figref{Fig:s15.0--Om0--B0--mshells}.

Core bounce occurs after 168 milliseconds of collapse; the shock forms
at a mass coordinate of $M_{\mathrm{sh}} \approx 0.45 ~ \msun$ and
starts propagating outwards.  A strong negative gradient of the
electron fraction triggers prompt PNS convection in a layer between an
inner radius of about 15 and an outer radius that is initially at
about 50 km and decreases to about 40 km at the end of the simulation.
In \figref{Fig:s15.0--Om0--B0--mshells}, we show the upper boundary of
the PNS convection zone by a solid black line.  Convective velocities
inside the PNS reach maximum values of $\approx 5 \times 10^8 ~ \cms$
and angle-averaged r.m.s.~values of $\approx 2.5 \times 10^8 ~ \cms$.

After roughly 50 ms post-bounce, the shock starts to oscillate as the
standing accretion shock instability
\citep[SASI][]{Blondin_Mezzacappa_DeMarino__2003__apj__Stability_of_Standing_Accretion_Shocks_with_an_Eye_toward_Core-Collapse_Supernovae}
develops in the post-shock region.  Perturbations created at the shock
wave are advected with the flow towards the PNS, leading to the
parallel alignment of the patterns of lateral velocity and of the
(grey) mass-shell lines lines in \figref{Fig:s15.0--Om0--B0--mshells}.
From these perturbations, the SASI develops in the post-shock region.
Typical transverse velocities associated with the SASI reach values
about 3 times as large as the convective velocities deeper inside the
PNS.  With increasing time, the SASI begins, starting in the immediate
post-shock region, to affect a larger fraction of the post-bounce
volume, until at $t \gtrsim 100~\mathrm{ms}$ the PNS convection zone
is separated from the SASI region by a thin stable layer located
roughly between densities of $10^{11}$ and $10^{12} ~ \gccm$.  The
border between the stable layer and the SASI region is shown in
\figref{Fig:s15.0--Om0--B0--mshells} by a solid black line.

For later reference, we list the definitions we use for the four
distinct regions of the core and their respective boundaries:
\begin{enumerate}
\item The \emph{inner hydrodynamically stable PNS} extends from the
  origin to roughly the radius where the density of the gas drops
  below $10^{14} \ \gccm$.
\item This inner core is surrounded by the \emph{PNS convection zone},
  characterised by a negative gradient of the electron fraction.  The
  outer boundary of this layer is associated with the location of the
  minimum of the $Y_e$ profile.
\item A \emph{stable layer} surrounds the PNS convection zone and
  separates the latter from the
\item \emph{gain region} where hot-bubble convection and the SASI
  operate.  The boundary between the stable and the gain layer is
  approximately given by the gain radius that defines the transition
  from neutrino cooling below to neutrino heating above.
\end{enumerate}

Despite growing SASI activity and increasing entropy in the post-shock
region and despite a slowly growing radius of the stalled shock, no
explosion takes place until 550 ms after bounce. This is in agreement
with simulations for 15$\msun$ stars (but different progenitor
models) by
\cite{Marek_Janka__2009__apj__Delayed_Neutrino-Driven_Supernova_Explosions_Aided_by_the_Standing_Accretion-Shock_Instability}
and
\cite{Burrows_et_al__2006__apj__A_New_Mechanism_for_Core-Collapse_Supernova_Explosions}.

\subsection{Models with weak initial fields}
\label{sSek:Res-Bw}

Models \emph{\modelname{s15-B10}} and \modelname{s15-B11} ($b_0 =
10^{10,11}~\mathrm{G}$, respectively) exhibit very similar dynamics
compared both to each other and to the non-magnetised reference model.
The structure of the core is during several hundred milliseconds of
post-bounce evolution the same as that of model \modelname{s15-B0}: a
convective layer inside the proto-neutron star at $\rho \gtrsim
10^{12} ~ \gccm$ and a SASI region, separated by a thin stable
deceleration layer; hence, the profiles of entropy and $v_{\theta}$
are in qualitative agreement with the ones shown in
\figref{Fig:s15.0--Om0--B0--mshells}.

\begin{figure*}
  \centering
  \includegraphics[width=17cm]{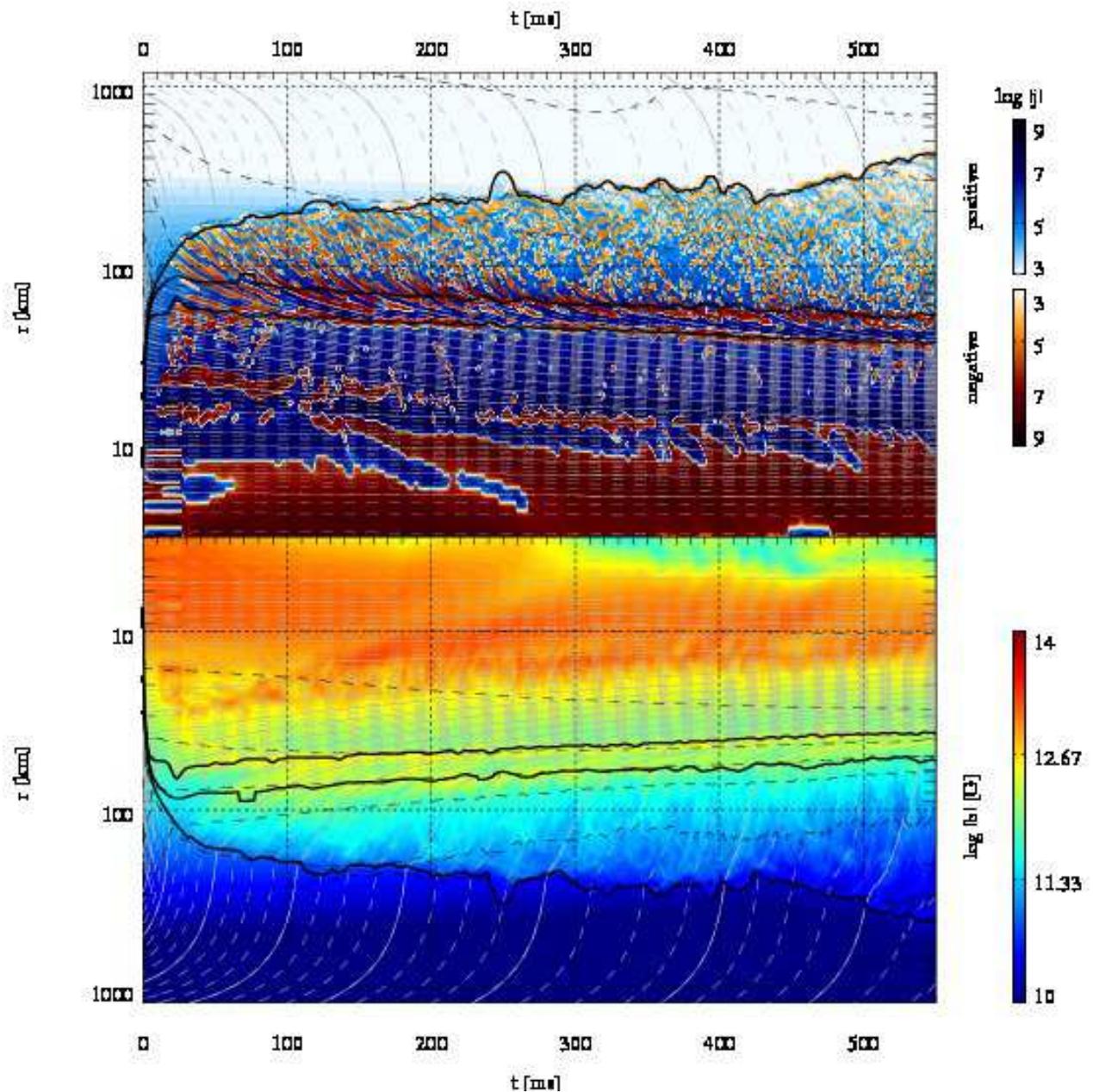}
  \caption{
    Evolution of model \modelname{s15-B10}.  The panels show in colour
    coding the logarithm of the lateral average of the current density
    (\textit{top panel}; the intensity of the colour encodes the
    absolute value and the blue and red colours distinguish between
    positive and negative signs) and the logarithm of the mean
    magnetic field strength (\subpanel{bottom panel}).  The lines have the
    same meaning as in \figref{Fig:s15.0--Om0--B0--mshells}.
  }
  \label{Fig:s15--B10--mshells1}
\end{figure*}

We discuss the evolution of the magnetic field of model
\modelname{s15-B10} in the following.  A space-time diagram of the
model (\figref{Fig:s15--B10--mshells1}) shows angular averages of the
current density\footnote{For simplicity, we omit the factor
  $\frac{4\pi}{c}$ in the definition of the current density and
  consider only the $\phi$-component, neglecting the vanishing
  components $j^r$ and $j^{\theta}$.} $j = (\vec \nabla \times \vec
b)^{\phi}$ and the field strength.  We use the current density to
detect spatial variations of the magnetic field.  While not all
patterns of $\vec j$ are necessarily caused by \Alfven~waves,
\Alfven~waves are associated with a non-zero current density.  Hence,
we can try to identify these waves in the current distribution.

Without changing the field geometry, compression amplifies the
magnetic field during collapse along with the maximum density as
$\rho_{\mathrm{max}}^{2/3}$, leading to a gain of a factor of $\approx
500$ and $\approx 25$ in the maximum field strength and total magnetic
energy, respectively.

\begin{figure}
  \centering
  \resizebox{\hsize}{!}{\includegraphics{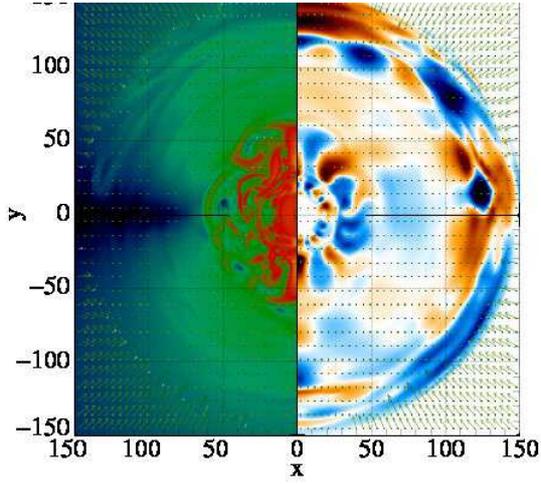}}
  \caption{
    Snapshot of model \modelname{s15-B10} shortly after the onset of
    PNS convection at $r \lesssim 50 ~ \km$.  The \subpanel{left} and
    \subpanel{right panels} show the logarithm of the magnetic field
    strength and the deviation of the electron fraction from its
    angular average.  Vectors indicate the velocity field.  }
  \label{Fig:s15.0--Om0--B10--conv-ampl}
\end{figure}

Further amplification occurs in the PNS convection zone (see
\figref{Fig:s15.0--Om0--B10--conv-ampl}).  Within roughly 10 ms after
the onset of PNS convection, the mean field is amplified by a factor
of $\sim 3$ between radii of $\sim 20$ and $\sim 60$ km.  Field growth
is driven by the convective overturn motions of the gas (indicated in
the deviations of $Y_e$ from its angular mean value, see \subpanel{right
  half} of \figref{Fig:s15.0--Om0--B10--conv-ampl}) and leads to a
filamentary field topology (same figure, \subpanel{left half}).  This
strongly non-uniform structure is reflected in the large fluctuations
of the current density visible in the top panel of
\figref{Fig:s15--B10--mshells1} (first $\sim$ 100 ms at radii between
$\sim 15$ and $\sim 60$ km).  After about 100 ms, the amplitude of
variations of the current density in the convective layer decreases.
While $\vec j$ still varies on short temporal and spatial scales,
larger coherent fluctuations are mostly absent during this phase:
compare the larger red ($j^{\phi} < 0$) patches in the PNS convection
zone at $t \lesssim 100 ~ \ms$ to the convection zone at later times
standing out as a broad blue ($j^{\phi} > 0$) band with little
contribution of negative currents.  In this phase, the angular average
of the current density in the PNS convection zone is very regular and
shows little temporal variation.  It is dominated by one large-scale
pattern, viz.~columns of strong magnetic field generated by flows
converging at the axis due to the constraint of axisymmetry
(\figref{Fig:s15.0--Om0--B10--conv-ampl}).  These very persistent
features are associated with a strong positive current density whose
value is dominating whatever fluctuations there may be due to
convective cells at the same radii.

As soon as the SASI develops, we find appreciable lateral velocities
also in the region above the PNS convection zone.  An early stage of
the development of the SASI can be seen already in the $Y_e$ variation
close to the shock in \figref{Fig:s15.0--Om0--B10--conv-ampl}: small
deformations of the shock excite perturbations advected inwards.
These perturbations are associated with kinks in the magnetic field
lines.  Geometrically, they take the form of sheets oriented parallel
to the shock, i.e., roughly at constant radius, containing a magnetic
field with a strong $\theta$-component.  This translates into a
non-zero derivative $\partial_{r} \, b^{\theta}$, i.e., a non-zero
current density.  Consequently, the perturbations can be identified
easily in the top panel of \figref{Fig:s15--B10--mshells1} as the
alternating blue and red bands starting at the shock and falling
towards the deceleration layer.  Propagating at the velocity of fluid
accretion, these bands are parallel to the mass shells (grey lines).

\begin{figure}
  \centering
  \resizebox{\hsize}{!}{\includegraphics{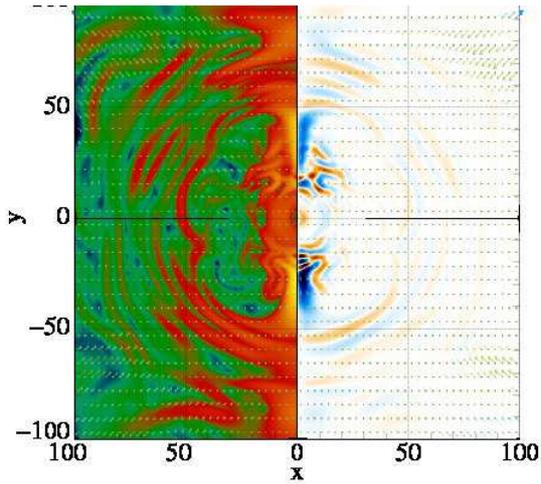}}
  \caption{
    Snapshot of model \modelname{s15-B10} at $t \approx 120~\ms$
    post-bounce.  The panels show, apart from the velocity field, the
    logarithm of the magnetic field strength (\subpanel{left}) and the
    current density (\subpanel{right}).  }
  \label{Fig:s15.0--Om0--B10--conv-colm}
\end{figure}

When the accretion flow decelerates above the PNS convection zone,
these sheets pile up at the deceleration layer
(cf.~\figref{Fig:s15.0--Om0--B10--conv-colm} at $r \gtrsim 50 ~ \km$),
leading to a local increase of the magnetic energy.  The bottom panel
of \figref{Fig:s15--B10--mshells1} indicates that this process happens
throughout the entire evolution and magnetic energy continues to
accumulate.

\begin{figure}
  \centering
  \resizebox{\hsize}{!}{\includegraphics{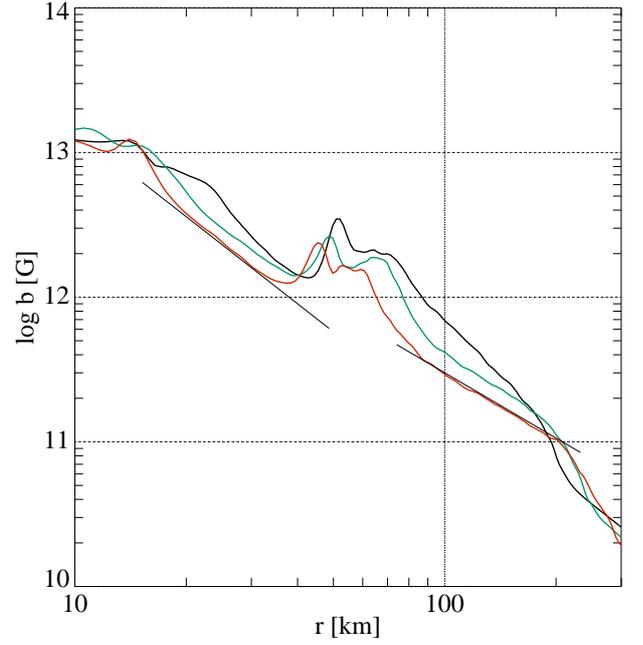}}
  \caption{
    Average radial profiles of the field strength of model
    \modelname{s15-B10} in the post-bounce phase (time averages over
    $t \in [100;200]~\ms$ (black), $t \in [200;300]~\ms$ (green), $t
    \in [300;400]~\ms$ (red) post-bounce).  The thin straight lines
    show profiles $\propto r^{-2}$ ($r < 50 \km$) and $\propto
    r^{-3/2}$ ( $r > 80~\km$).  }
  \label{Fig:s15--B10--avge-b-profile}
\end{figure}

\subsubsection{Sub- and superalfv\'enic regions}
\label{ssSek:subsupreg}

In the following, we discuss the propagation of MHD waves in the core.
We will not touch upon fast modes and focus only on slow and
\Alfven~waves.  Whereas \Alfven~and slow modes propagating across the
field lines are clearly distinct solutions (they have, e.g., different
polarisation states and different propagation speeds), they degenerate
when propagating along the field lines.  In this case, they have
identical propagation speeds, viz.~the \Alfven~speed.  Because we
expect the dynamics to be dominated by the modes propagating fastest,
i.e., the ones travelling (anti-)parallel to the field lines, we will
consider these modes only.  Because of the degeneracy of slow and
\Alfven~modes in such a case, we will refer to these modes as
\Alfven~waves for short, postponing an investigation of the possible
differences between the two classes of modes.  It is expected
(J.~Guilet, \textit{private communication}) that slow modes behave
very similarly to \Alfven~waves.

An essential requisite for this evolution is that the flow is
superalfv\'enic, i.e., the flow velocity is larger than the
\Alfven~velocity of the magnetic field
\begin{equation}
  \label{Gl:Res--cA}
  c_{\mathrm{A}} = \frac{\sqrt{\vec b^2}}{\sqrt{\rho}} 
  = 
  10^6 \, \cms ~
  \left( \frac{|\vec b|}{ 10^{12} \, \mathrm{G}} \right)
  \left( \frac{ 10^{12} \, \gccm}{ \rho} \right) ^ {1/2}
  .
\end{equation}
A magnetic field $\vec b$ allows for the propagation of \Alfven~waves
in and against the direction of the field at the same propagation
speed, $c_{\mathrm{A}}$ relative to the flow of the gas carrying the
magnetic field, i.e., with a velocity $\pm \vec c_{\mathrm{A}} = \pm
\vec b / |\vec b| \, c_{\mathrm{A}}$.  The total velocity of such a
pair of waves in the lab frame, where the gas moves with a velocity
$\vec v$, is $\vec v \pm \vec c_{\mathrm{A}}$.  Hence, an
\Alfven~point develops where the component of the \Alfven~speed
directed along the fluid velocity has the same absolute value as $\vec
v$.  At that point, one of the pair of \Alfven~waves, viz.~the one
with an \Alfven~velocity antiparallel to $\vec v$, is trapped , i.e.,
has vanishing total velocity, while the other one passes the
\Alfven~point at a total speed $2 |\vec v|$.

\begin{figure*}
  \centering
  \includegraphics[width=17cm]{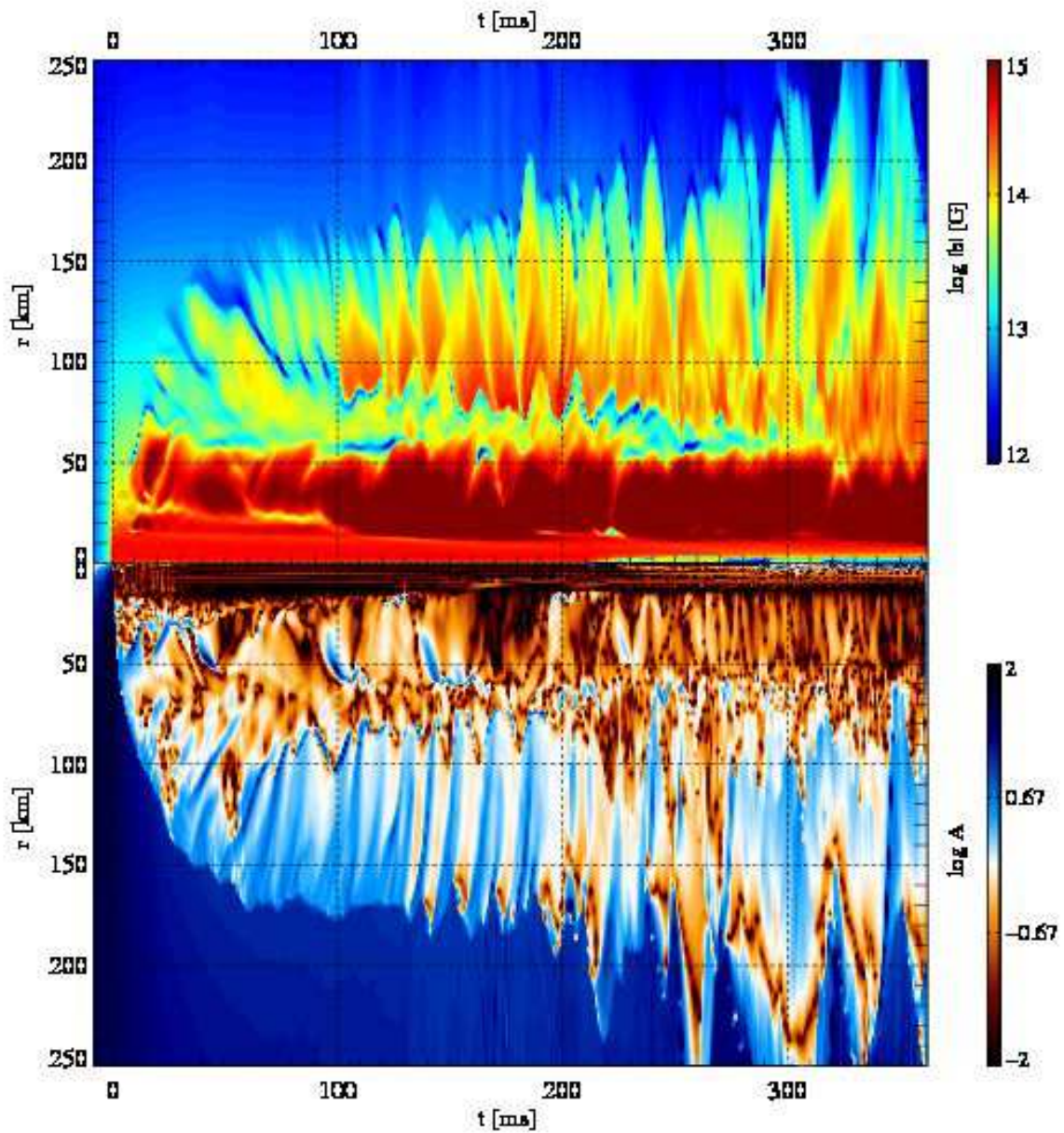}
  \caption{
    Space-time diagram of the magnetic field strength (\subpanel{top
      panel}) and the logarithm of the \Alfven~number (\subpanel{bottom
      panel}) of model \modelname{s15-B11.5} along the north pole.
  }
  \label{Fig:s15--B11.5--b2-pol}
\end{figure*}

To quantify the importance of the magnetic field, we define the total,
parallel, and perpendicular \Alfven~numbers,
\begin{eqnarray}
  \label{Gl:Res--Altot}
  \Alnum_{\mathrm{tot}}
  & = & \frac{|\vec v|}{|\vec c_{\mathrm{A}}|},
  \\
  \label{Gl:Res--Alpar}
  \Alnum_{\parallel}
  & = &
  \frac{\vec v \cdot \vec c_{\mathrm{A}}}{\vec c_{\mathrm{A}}^2},
  \\
  \Alnum_{\perp}
  & = & 
  \frac{|\vec v \times \vec c_{\mathrm{A}}|}{\vec c_{\mathrm{A}}^2}.
  \label{Gl:Res--Alprp}
\end{eqnarray}
While $\Alnum_{\mathrm{tot}}$ compares kinetic and magnetic energies
directly, $\Alnum_{\parallel}$ and $\Alnum_{\perp}$ measure two
effects important to \Alfven~waves propagating along the field:
\begin{itemize}
\item $\Alnum_{\parallel}$ regulates the total (lab-frame) propagation
  speed of the wave, composed of the (comoving) \Alfven~velocity and
  the flow speed, $\pm \vec c_{\mathrm{A}} + \vec v$;
  $|\Alnum_{\parallel}| = 1$ indicates an \Alfven~point where the
  waves stall.
\item $\Alnum_{\perp}$ is a parameter relevant to the excitation of
  perturbations orthogonal to the magnetic field, comparing the
  velocity displacing magnetic field lines with the tension of the
  field lines.  \Alfven~waves rely on magnetic tension as restoring
  force; hence, their excitation occurs for $\Alnum_{\perp} < 1$;
  otherwise, the field can be distorted (e.g., amplified in a
  stretch-twist-fold dynamo), but the distortions are beyond the
  linear regime of \Alfven~waves.
\end{itemize}

The gas flows triggered by PNS convection as well as by the SASI are
highly superalfv\'enic.  Though there is, in principle, a small stable
region between the convection and the SASI regions, where the
accretion is decelerated and only small lateral velocities are
present, the flow in this region remains superalfv\'enic because of
overshooting from the PNS convection and the SASI regions.  Travelling
along a field line connecting the convection zone with the SASI
region, one observes a drop of all three \Alfven~numbers in the
deceleration region, but they still remain considerably above unity.
Hence, there is neither any significant excitation of \Alfven~waves
nor is there an \Alfven~point where amplification as proposed by
\cite{Guilet_et_al__2010__ArXive-prints__Dynamics_of_an_Alfven_surface_in_CCSNe}
could happen.  Magnetic perturbations are unable to travel upwards
against the accretion flow; furthermore, the relative weakness of the
field prohibits dynamic back-reaction.  Thus, this is a purely
kinematic effect.

The combination of these effects leads to the time evolution of the
average field profile shown in \figref{Fig:s15--B10--avge-b-profile}.
Across the PNS convection zone, the average field strength drops by an
order of magnitude.  At its upper boundary, the pile-up of field
accreted from the outer layers leads to a strong local increase of the
field strength, while in the SASI region, the radial decrease of the
field continues following a slightly flatter law ($b \propto
r^{-3/2}$) than in the PNS convection zone ($b \propto r^{-2}$).  The
steepening of the slope at $r \sim 200 ~ \km$ marks the minimum radius
of the shock wave.

For model \modelname{s15-B10}, the flow is superalfv\'enic almost
everywhere.  Sizeable subalfv\'enic regions develop only for
significantly stronger fields.  The convective flows of model
\modelname{s15-B11.5}, with an initial field strength given by $b_0 =
10^{11.5} ~ \mathrm{G} \approx 3.16 \times 10^{11} ~ \mathrm{G}$, are
only moderately superalfv\'enic.  In particular, along the pole, where
the magnetic field is enhanced artificially for symmetry reasons, the
\Alfven~numbers are below unity (\figref{Fig:s15--B11.5--b2-pol}).
The thin layer separating PNS convection and hot bubble consists of
both sub- and superalfv\'enic regions, while the accretion flow is
mostly superalfv\'enic.

In analogy to the model discussed above, perturbations of the magnetic
field created at the shock wave are advected towards the PNS.  The
field piling up above the PNS, dominated by a strong lateral component
generated by the complex overturns of the gas flow, forms shells
around the PNS.  Since the radial component is weak, \Alfven~waves are
bound to travel at roughly constant radius.  Consequently, the
location of an \Alfven~point along a field line is determined not by
the radial flow alone but mostly by the lateral velocity.  Whereas the
former tends to go through zero in the deceleration region, the latter
can, in principle, have a stochastic and chaotic behaviour and
dynamical evolution.

\begin{figure}
  \centering
  \resizebox{\hsize}{!}{\includegraphics{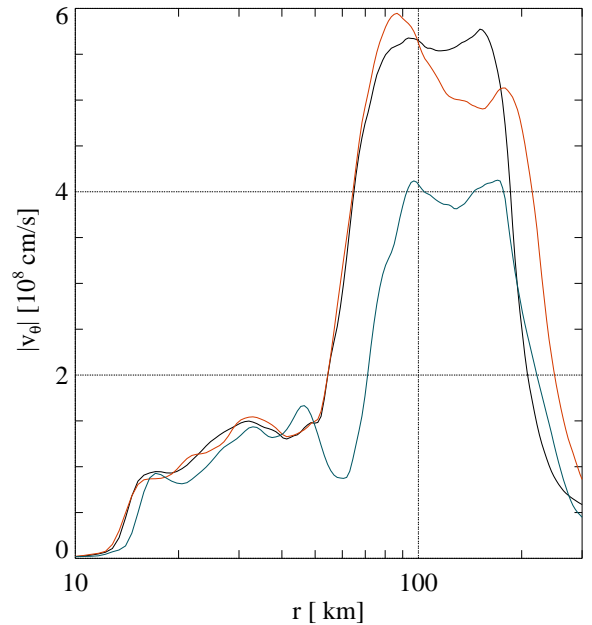}}
  \caption{
    Root-mean-square averages of the $\theta$-component of the
    velocity as a function of radius for models \modelname{s15-B10}
    (black), \modelname{s15-B11} (red), and \modelname{s15-B11.5}
    (blue).  Time averages over several tens of milliseconds in the
    late post-bounce phase (at $t \approx 300 \ \ms$) are shown.  }
  \label{Fig:s15--B10-B11-B11.5--accu-vgl}
\end{figure}

Thus, while there may be \Alfven~points in this region, their location
is highly variable; they may even, depending on the flow, cease to
exist for a certain time when the flow becomes superalfv\'enic or
subalfv\'enic in the entire region.  This may have adverse effects on
the efficiency of the amplification process proposed by
\cite{Guilet_et_al__2010__ArXive-prints__Dynamics_of_an_Alfven_surface_in_CCSNe}.
Despite the stochastic nature of these variations, we find a
systematic trend: the stronger the magnetic field is, the farther out
stretches the layer of accumulated magnetic field on average.  Since
this layer corresponds to the hydrodynamically stable region between
PNS convection and the SASI region, the trend of a growing radius of
the region where the field accumulates can be seen most easily in the
average profiles of the lateral velocity shown in
\figref{Fig:s15--B10-B11-B11.5--accu-vgl}.  This effect is an, albeit
small, indication of dynamic back-reaction of the field onto the flow.

\subsubsection{Accretion columns}
\label{ssSek:Acccol}

The accretion columns display much stronger radial components of the
field and the velocity, while non-radial fields of a considerable
strength may be present in the flux sheets surrounding them.
Accretion proceeds mostly through a narrow stream at the poles, and
mass conservation leads to an increase of the flow speed as the gas is
accreted through a small solid angle rather than uniformly over all
angles.  Eventually, the flow reaches sound speed.  Analogously, flux
conservation increases the magnetic field strength in the accretion
funnels.  Depending on the initial field strength, the \Alfven~number
of the accretion flow and $\beta = P / P_{\mathrm{mag}}$ may be of
order unity or much greater.

Due to the high Mach number, pressure perturbations of the accretion
flow may steepen, in some cases yielding discontinuities.  In
particular, weak shock waves may mark the lower end of the column,
providing a very abrupt deceleration of the gas.  Due to the
dissipation of kinetic into internal energy, the entropy of the matter
above the shock is higher than that of the gas in the deceleration and
PNS convection layers where $S$ is nearly uniform.

In the presence of a magnetic field, these features are not purely
hydrodynamic but rather MHD structures associated with a bending of
the field lines, i.e.~a current density: the field, predominantly
radial in the accretion column, is folded by the relatively high
lateral velocity in the deceleration region and is mostly mostly
lateral in this layer.  This effect is weakest in the polar accretion
funnels because the axial symmetry restricts the development of a
non-radial field considerably.  In all accretion columns, both
velocity and magnetic field are dominated by their respective radial
components; for the polar accretion flow, this dominance of the radial
components holds even at radii below the lower end of the accretion
stream, i.e., for $r \lesssim 70-100 ~ \km$.  Accretion streams at
intermediate latitudes tend to punch into gas threaded by strong
lateral field lines and moving at considerable speed along constant
radius.

For a sufficiently strong magnetic field, an \Alfven~point may be
located in the accretion stream; for a steep gradient of $v_r$ or even
a discontinuity at the lower end of the stream, this is a most natural
place for the \Alfven~point.

Conditions are most favourable for the formation of a stable
\Alfven~point in radial polar accretion streams.  We show the time
evolution of the magnetic field and the \Alfven~number along the north
pole of model \modelname{s15-B11.5} in
\figref{Fig:s15--B11.5--b2-pol}.  Initially, accretion is roughly
spherical, but after $t \approx 80 ~ \mathrm{ms}$, polar accretion
streams begin to form; as a consequence, the radial velocity behind
the shock increases abruptly.  This stream remains active for the rest
of the simulation.  It is, however, subject to strong fluctuations on
time scales of about 10 milliseconds.  These oscillations affect the
outer and inner termination radius of the accretion flow, the
velocity, and the magnetic field.  They are caused by lateral motions
due to convection and the SASI in the surrounding gas that lead to an
episodically varying cross section of the funnel.  Following the
variations of the accretion velocity, the Mach number of the flow
oscillates around unity.  For this model, the same holds for the
\Alfven~number.

The accretion flow terminates in a steep gradient hosting an
\Alfven~point.  This point is first, at $t \approx 100 ~ \ms$
post-bounce, situated at a radius of 80 kilometres, i.e., about 30
kilometres above the PNS convection layer; later it retreats to lower
radii as the core contracts.  The intermediate layer between the
termination point of the accretion column and the PNS convection layer
is, though threaded by a weaker magnetic field, filled by gas at
subalfv\'enic velocities.  Hence, at these radii, \Alfven~waves could
propagate from below towards the accretion flow.  Given a typical
\Alfven~velocity in this layer of $c_{\mathrm{A}} \sim 2 \times 10^{8}
~ \cms (b_0/ 10^{11}~\mathrm{G})$, they should traverse the
intermediate layer within $\tau_{\mathrm{A}} \sim 10 ~ \mathrm{ms} /
(b_0/ 10^{11}~\mathrm{G})$, i.e., on the time scale of the
oscillations of the accretion column.  This has two consequences:
first, it complicates the identification of \Alfven~waves, and,
second, it may limit the applicability of the scenario of
amplification of \Alfven~waves in this context.

The fluctuations of the accretion velocity and the position (radius,
possibly also latitude) of the termination point of the accretion
column induce perturbations propagating in the layer between the
accretion column and the PNS convection region, i.e., outside a radius
of about 50 kilometres and inside a radius that recedes gradually from
about $80-100$ kilometres at $t \approx 100 ~ \ms$ to roughly $60 ~
\mathrm{km}$ at $t \approx 300 ~ \ms$.  Partially, these perturbations
can be entrained in matter flowing upwards and thus sustaining the
magnetic field in the hot-bubble region.  Additionally, overshoot from
the convective layer below contributes to a certain degree to
maintaining the turbulence in the deceleration region.

\subsubsection{Field amplification}
\label{ssSek:Fampl}

\begin{figure}
  \centering
  \resizebox{0.925\hsize}{!}{\includegraphics{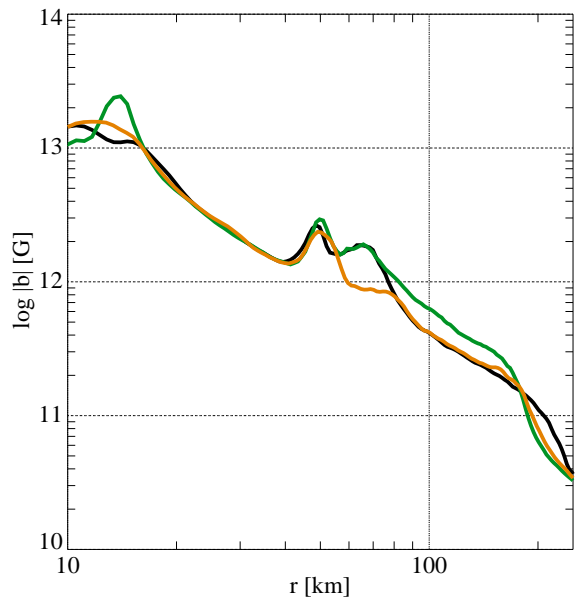}}
  \caption{
    Space-time averages of the magnetic field strength of models
    \modelname{s15-B10} (black), \modelname{s15-B11} (green), and
    \modelname{s15-B11.5} (orange line).  We average over times
    between $t = 200 ~ \ms$ and $t = 300 ~ \ms$ post-bounce.  The
    fields of models \modelname{s15-B11} and \modelname{s15-B11.5} are
    scaled according to their initial strength.  }
  \label{Fig:s15.0--Om0--B10-B11.5--bulk-b2}
\end{figure}

\begin{figure}
  \centering
  \resizebox{0.925\hsize}{!}{\includegraphics{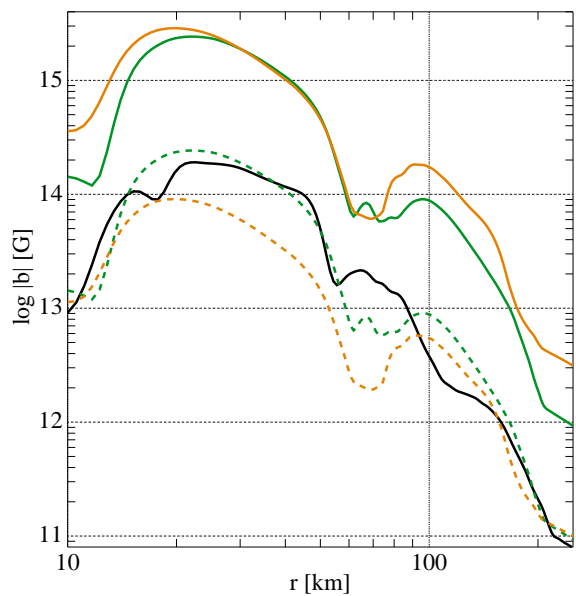}}
  \caption{
    \subpanel{Left panel:} Temporally averaged (over post-bounce times of
    $120 \mathrm{ms} \lesssim t \lesssim 200 ~ \mathrm{ms}$) radial
    profiles of the magnetic field strength along the north pole for
    models \modelname{s15-B10} (black), \modelname{s15-B11} (green),
    \modelname{s15-B11.5} (orange).  The field strength is represented
    by solid lines; the dashed lines show the field strength scaled by
    the ratio of initial field strength.
  }
  \label{Fig:s15--B10-B11-B11.5--b2-pol-avge}
\end{figure}

Though, in principle, the dynamics in this layer can be expressed in
terms of basic MHD waves, it is not trivial to identify these simple
solutions in the highly variable fields, preventing deeper insights
into the field amplification mechanisms.  Therefore, we must try to
find indirect evidence for the amplification processes at work.  To
this end, we distinguish between kinematic and dynamic amplification,
i.e., processes that do not depend on the seed magnetic field (at
least during an initial phase when it is still sufficiently weak) and
processes for which the amplification is some function of the seed
field, respectively.

Convection and the SASI amplify the field kinematically.  Hence, they
should amplify the field by a given factor independent of the seed
field.  This is demonstrated by the comparison of space-time averages
of radial profiles of the magnetic field strength of models with
different initial fields shown in
\figref{Fig:s15.0--Om0--B10-B11.5--bulk-b2}. Corrected for the scaling
with the initial field strength, the profiles of all three models are
remarkably similar.  The main differences appear in the deceleration
region, where neither the radial collapse nor convection or the SASI
are the main amplification mechanisms.

\begin{figure*}
  \sidecaption
  \includegraphics[width=6cm]{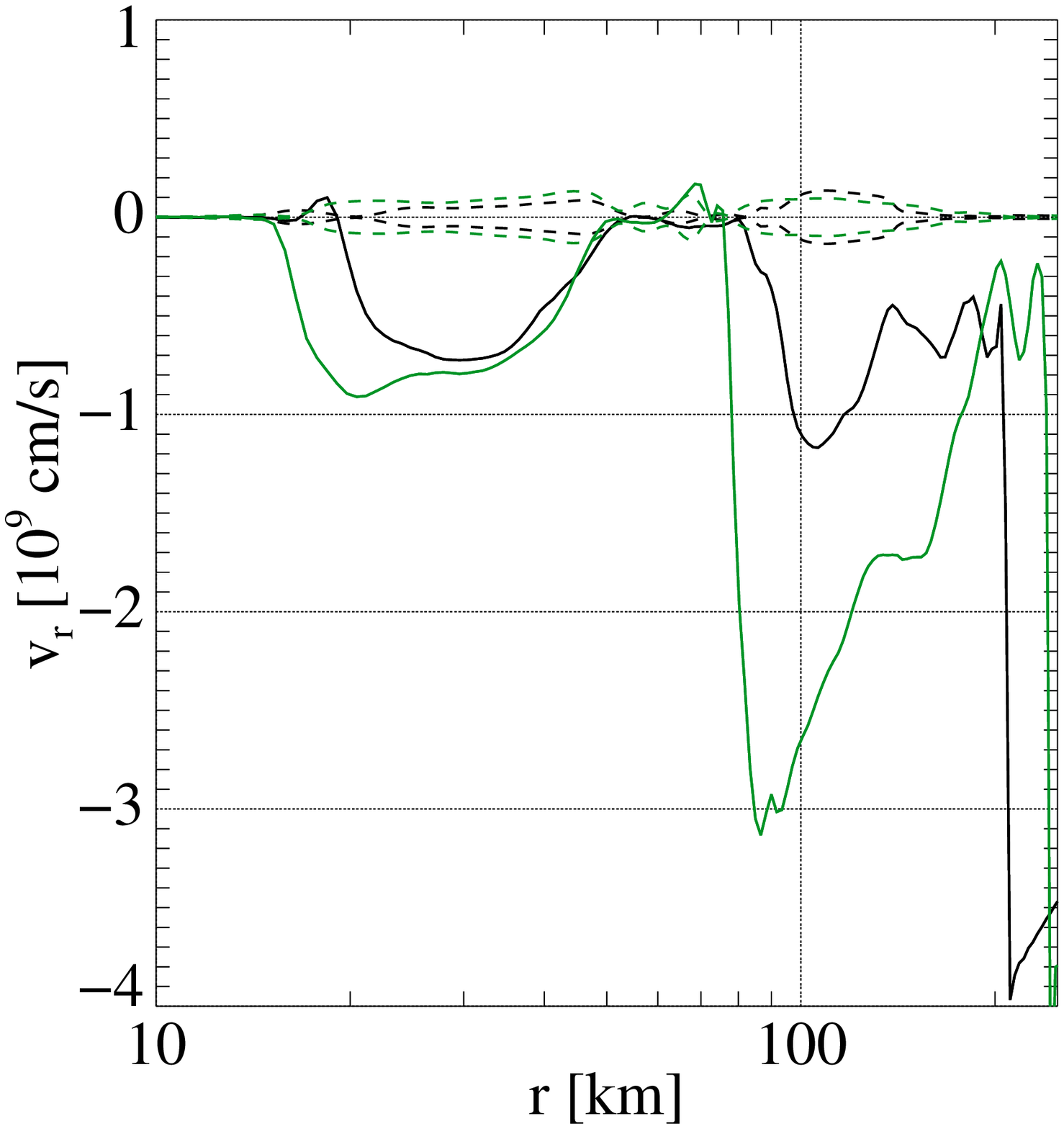}
  \includegraphics[width=6cm]{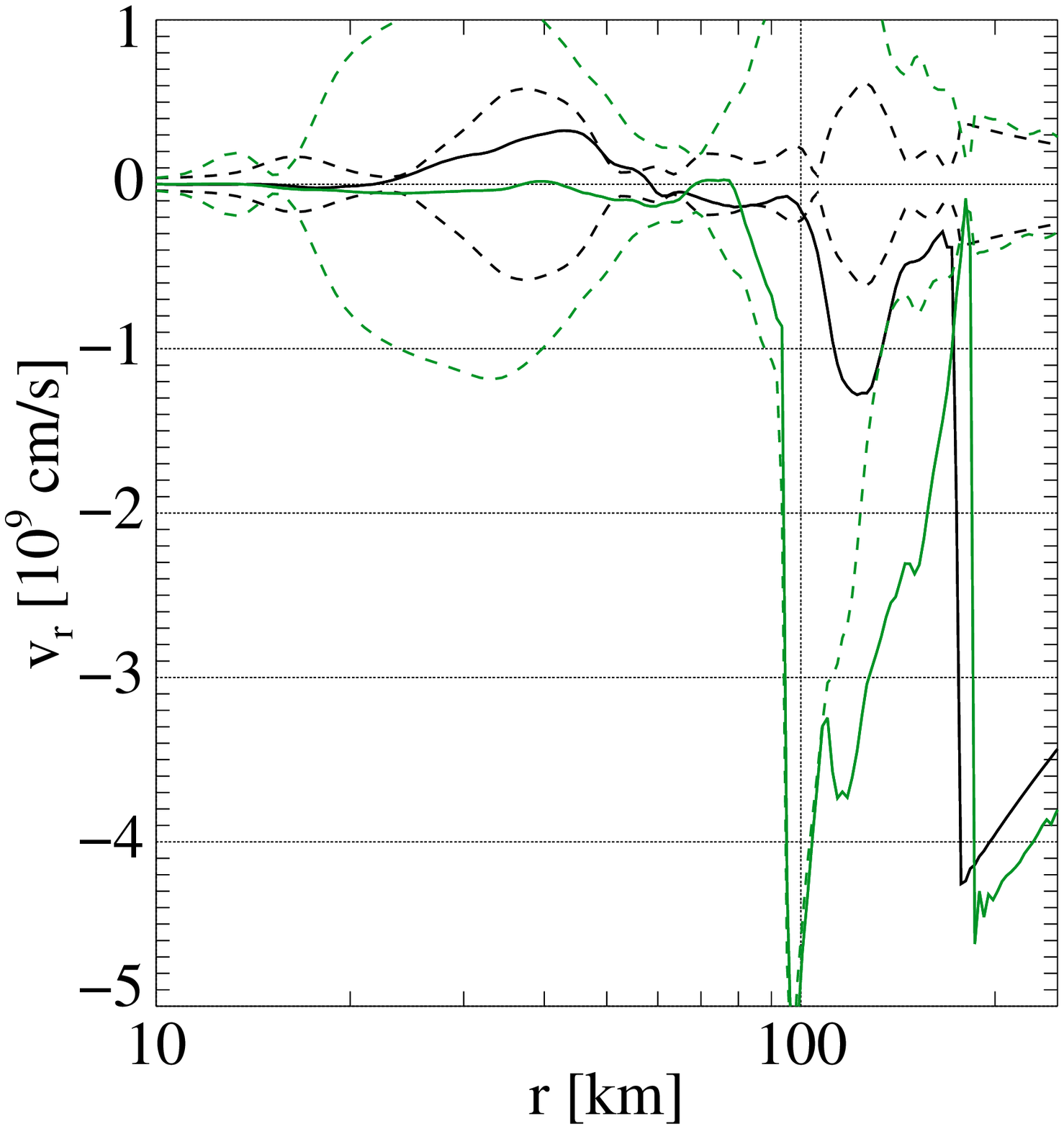}
  \caption{
    The \subpanel{top panels} show profiles of the radial velocity
    (solid lines) along the north pole of models \modelname{s15-B10}
    (\subpanel{left}) and \modelname{s15-B11.5} (\subpanel{right});
    the dashed lines show the \Alfven~velocity, $c_{\mathrm{A}}$, and
    its negative, $- c_{\mathrm{A}}$.  The bottom panels depict
    profiles of the magnetic field strength of the same models.  In
    all panels, black and green lines denote the profiles at $t
    \approx 100 ~ \mathrm{ms}$ and $t \approx 200 ~ \mathrm{ms}$ post
    bounce, respectively.  }
  \label{Fig:s15.0--Om0--B10-B11.5--npol-vb}
  \includegraphics[width=6cm]{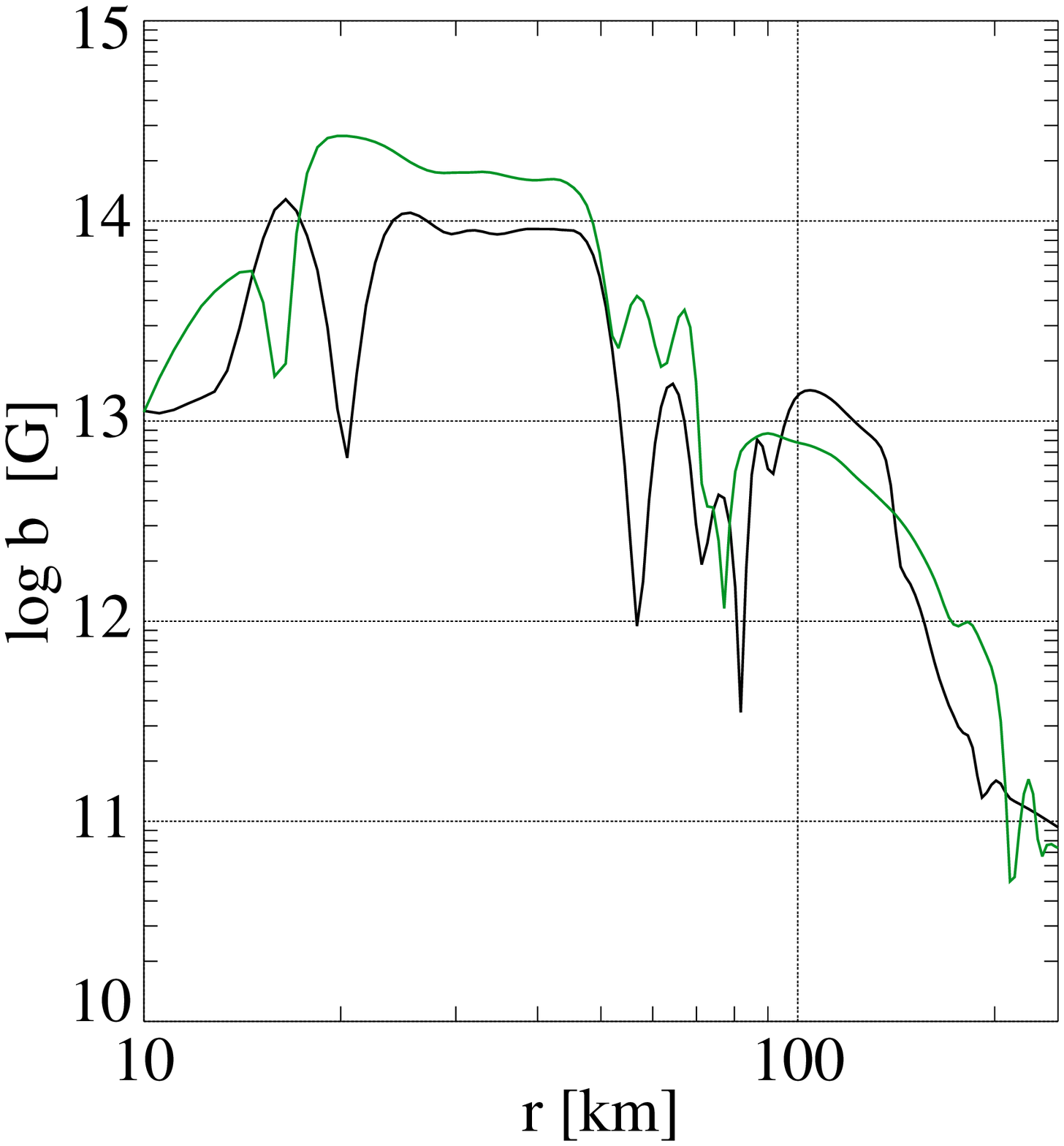}
  \includegraphics[width=6cm]{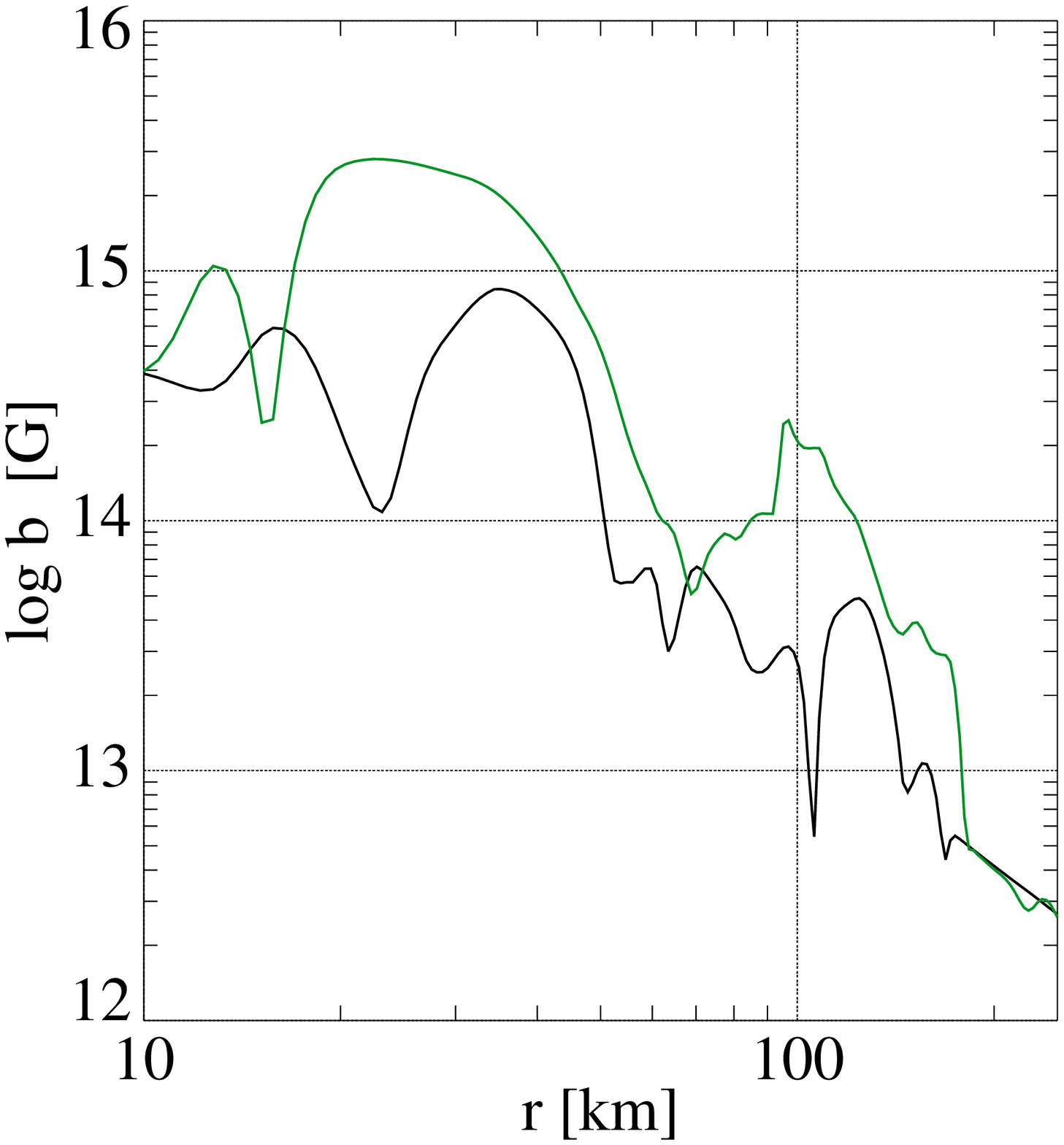}
\end{figure*}

The \Alfven~mechanism due to
\cite{Guilet_et_al__2010__ArXive-prints__Dynamics_of_an_Alfven_surface_in_CCSNe},
on the other hand, depends on the presence and the location of
\Alfven~points, hence on the initial field strength.  Furthermore, the
\Alfven-wave mechanism should be most efficient along the polar axis
(as discussed above and shown in \figref{Fig:s15--B11.5--b2-pol}).  A
comparison of the polar regions of models with different initial field
strengths should therefore allow us to determine the importance of the
\Alfven-wave mechanism.

We compare time averages of the field strength along the north pole
for different initial fields in
\figref{Fig:s15--B10-B11-B11.5--b2-pol-avge} and profiles for two
times in \figref{Fig:s15.0--Om0--B10-B11.5--npol-vb}.  In the PNS
convection zone, the field is amplified roughly by the same factor
irrespective of $b_0$, but only until it reaches kinetic
equipartition.  This limit is reached by model \modelname{s15-B11.5}.
The field decreases with radius in the convection zone and it exhibits
a further drop by about an order of magnitude at the upper boundary of
the convective layer inside the PNS.

After a few tens of milliseconds, the evolution of models
\modelname{s15-B10} and \modelname{s15-B11.5} diverges.  The
weak-field model shows a coherent stable downflow inside the
convection region (at radii between 20 and 50 kilometres) and a rather
smooth velocity profile in the surrounding layers (at radii above 70
kilometres; see the black line in the \subpanel{upper left panel} of
\figref{Fig:s15.0--Om0--B10-B11.5--npol-vb}).  \Alfven~points may be
located at the border between the PNS convection region and the
accretion flow.  The accretion flow transports the magnetic field
towards the convection zone and establishes a temporally averaged
profile varying around $|\vec v| \propto r^{-3}$ between the
convection zone and the shock wave.

In model \modelname{s15-B11.5} (for profiles at different times, see
the \subpanel{right panels} of
\figref{Fig:s15.0--Om0--B10-B11.5--npol-vb}), strong perturbations
originating in the convection zone propagate upwards against the
accretion flow (note the large positive velocity below $r \approx 50 ~
\km$).  Under the influence of perturbations converging from below and
above, the lower end of the accretion column develops an increasingly
steep profile until a discontinuity is formed a few tens of kilometres
above the PNS convection zone (visible in the green line at $r \approx
100 ~ \km$; the discontinuity located at $r \approx 200 ~ \km$ is the
stalled supernova shock wave).  Magnetic energy is transported towards
this point from both sides (some of the accreted field can be
identified in the black curve in the \subpanel{lower right panel} above $r
\approx 100 ~ \mathrm{km}$) and a region of enhanced field correlated
with the termination of the accretion flow is created, visible right
at the discontinuity (at $r \gtrsim 90 ~ \km$ in the green line).

Thus, upflows travelling from the PNS convection zone towards the
polar accretion columns play a role in stopping the accretion flow and
shaping the radial profile at its lower end.  We find that the lower
end of the equatorial accretion column develops steeper profiles, too,
but this is caused by a different effect: the accretion flow continues
to smaller radii until it is terminated at a radius of $r \approx 60 ~
\km$, i.e., it terminates when it hits the high-density layer near the
outer boundary of the PNS convection zone.  These discontinuities do
typically not coincide with an \Alfven~point and are not associated
with an enhancement of the magnetic energy.

Hence, we find that field amplification, while dominated by kinematic
effects such as convection and the SASI, has also a certain
contribution from the growth of the field by the superposition of
perturbations propagating in opposite directions between the regions
of PNS convection and the SASI.

\subsection{Stronger magnetic fields}
\label{sSek:Res-Bs}

For an initial field of $b_0 = 10^{12} ~ \mathrm{G}$ (model
\modelname{s15-B12}), we find, much like in the models with weaker
fields, a convection layer inside the PNS and, outside of the PNS, the
hot-bubble region dominated by the SASI.  Due to the much stronger
initial field, however, the accretion flow onto the PNS is
subalfv\'enic almost out to the shock wave.

\begin{figure}
  \centering
  \resizebox{\hsize}{!}{\includegraphics{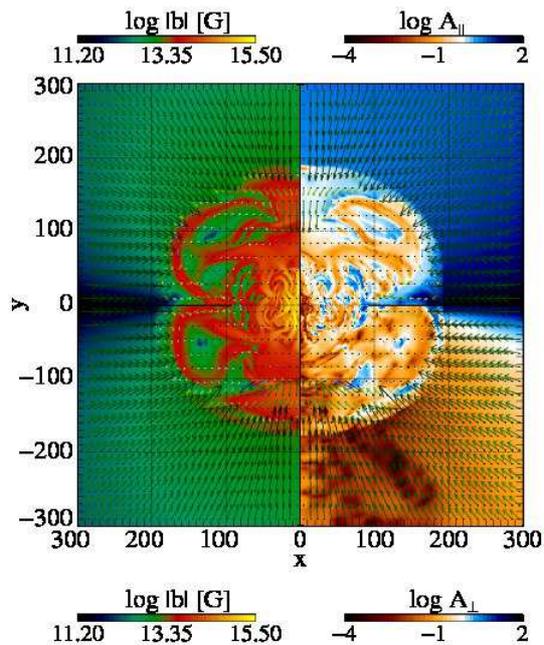}}
  \caption{
    Snapshot of model \modelname{s15-B12} at $t \sim 130$ ms
    post-bounce.  described roughly, and the \subpanel{top} and
    \subpanel{bottom right halves} show the logarithm of the parallel
    and perpendicular \Alfven~numbers, respectively.  Sub- and
    superalfv\'enic regions appear red and blue, respectively.  }
  \label{Fig:s15--B12--3000-bj}
\end{figure}

\begin{figure}
  \centering
  \resizebox{\hsize}{!}{\includegraphics{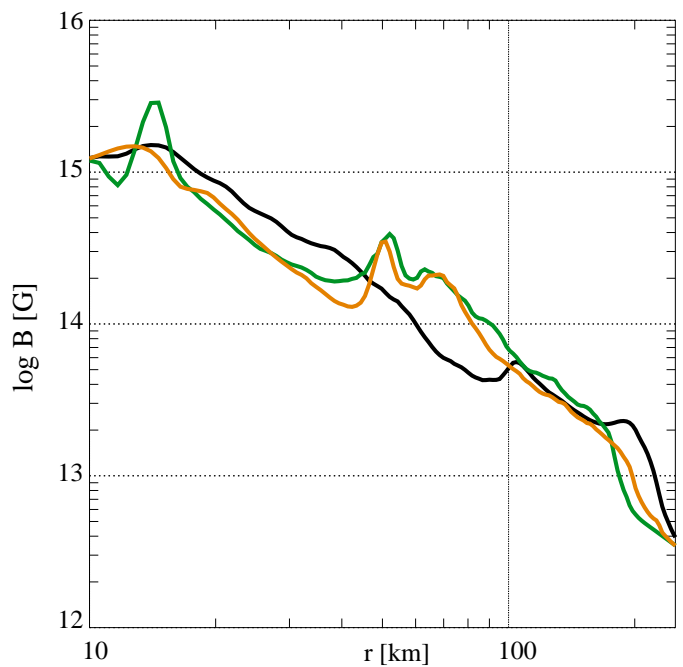}}
  \caption{
    Average (over angle and time, $160 ~ \mathrm{ms} \lesssim t
    \lesssim 220 ~ \mathrm{ms}$ post-bounce) profiles of the magnetic
    field strength of models \modelname{s15-B12} (black),
    \modelname{s15-B11} (green), and \modelname{s15-B10} (orange),
    corrected for the scaling of the initial field strength in the
    case of the latter two models.  }
  \label{Fig:s15.0--Om0--B10-B11-B12--bulk-b2}
\end{figure}

\begin{figure}
  \centering
  \resizebox{\hsize}{!}{\includegraphics{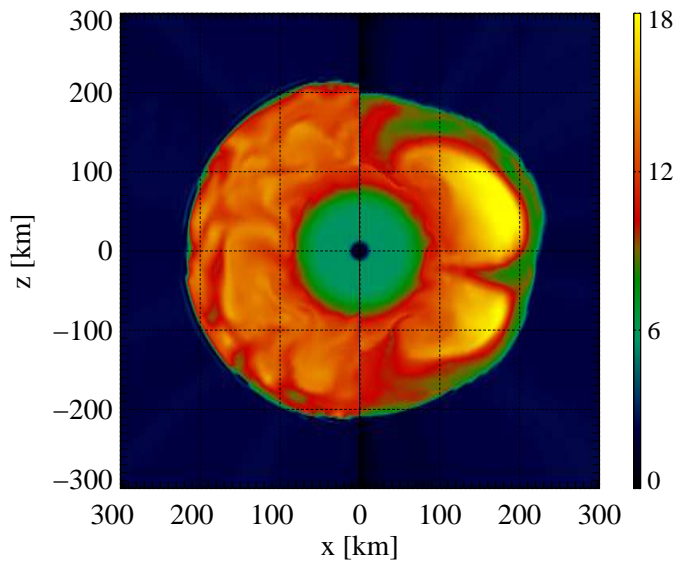}}
  \caption{
    Comparison of the entropy per baryon of models \modelname{s15-B10}
    (\subpanel{left half}) and \modelname{s15-B12} (\subpanel{right}) at $t
    \approx 170 ~ \ms$.
  }
  \label{Fig:s15.0--Om0--B1012--entvgl}
\end{figure}

\begin{figure}
  \centering
  \resizebox{\hsize}{!}{\includegraphics{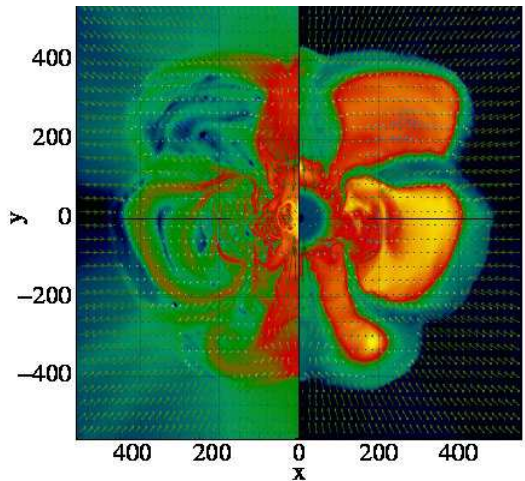}}
  \caption{
    Snapshot of model \modelname{s15-B12} at $t \approx 300 ~ \ms$.
    The \subpanel{left} and \subpanel{right halves} shows the
    logarithm of the magnetic field strength and the entropy per
    baryon, respectively.  }
  \label{Fig:s15.0--Om0--B12--late2d}
\end{figure}

The spatial structure of the model at $t \approx 130 ~ \ms$ is
displayed in \figref{Fig:s15--B12--3000-bj}.  The hot-bubble region is
dominated by a super-equipartition (w.r.t.~the velocity) magnetic
field; this is particularly evident in the large-scale arcs of the
magnetic field right behind the shock wave.  In these arcs, which
separate gas of a high entropy on the inside from low-entropy gas on
the outside, the magnetic field reaches equipartition with the
internal energy of the gas.  Gas falling through the shock wave is
deflected sideways towards two polar and an equatorial accretion
column, yielding a quadrupolar structure of the flow and the magnetic
field, in contrast to models with weaker fields, where the flow
structures are dominated by higher multipolarity.

Accretion proceeds along a few directions, viz.~in these columns, at
trans- and superalfv\'enic velocities, whereas the gas moves
subalfv\'enically between these accretion flows (the blue regions in
the \subpanel{right half} of \figref{Fig:s15--B12--3000-bj}).  In
these regions, the \Alfven~points lie far out in the accretion/SASI
region at more than 150 kilometres, whereas they are located roughly
at the deceleration radius in the downflows, i.e., at radii around 100
kilometres (see the borders between blue and red regions in the
\subpanel{right halves} of \figref{Fig:s15--B12--3000-bj}).  While
these regions exhibit some variations of their shape with time, they
are present during the entire post-bounce time.  It is also worth
mentioning that $\Alnum_{\perp}$ is less than unity in most of the
post-shock regions, meaning that perturbations of the magnetic field
can indeed be described in terms of \Alfven~waves, whereas
$\Alnum_{\parallel}$ varies between values greater and less than
unity, i.e., the waves are either trapped by the flow or can escape.

Due to the stable quadrupolar structure of the flow, the dynamics of
an accretion column, as outlined in \secref{ssSek:Acccol}, is
important for the evolution of this model.  Magnetic field is accreted
in the downflows, while it can rise against the accretion in the broad
regions of slowly moving gas between.  Similarly to model
\modelname{s15-B11.5}, waves travel upwards along the polar axis from
the PNS convection zone.  The accretion flow is decelerated abruptly
at a discontinuity at the lower end of the accretion column, and an
\Alfven~point is located there.  At this point, we find a pronounced
enhancement of the magnetic field.

\figref{Fig:s15.0--Om0--B10-B11-B12--bulk-b2} displays a comparison of
average profiles of the magnetic fields for models models
\modelname{s15-B12}, \modelname{s15-B11}, and \modelname{s15-B10},
corrected for the scaling of the initial field strength.  While the
models with weaker initial fields (green and orange lines) show a
distinct increase of the field strength at the outer boundary of the
PNS convection zone, i.e., at $r \approx 50 ~ \km$, a similar feature
is present for the model with stronger field at a much larger radius
of $r \sim 100 ~ \km$; in this case, the profile is described roughly
by a uniform power law out to this point.  The maximum field strength
of model \modelname{s15-B12} is of the order of $10^{15} ~ \mathrm{G}$
at the base of the convection zone.

Between the accretion columns, strong magnetic flux sheets are located
at the lateral and upper boundaries of the upwards flowing plumes.
This is visible, e.g., in the large arcs of field connecting the
accretion columns immediately behind the shock (see
\figref{Fig:s15--B12--3000-bj}).  Gas and magnetic field accreted in
the accretion columns rises in these regions from the deceleration
layer.  On the other hand, rising magnetic flux can, upon reaching the
boundaries of the plumes, be entrained in the accretion flow and thus
increase the accreted magnetic energy.

In \figref{Fig:s15.0--Om0--B1012--entvgl}, we compare the entropy
distribution of models \modelname{s15-B10} (\subpanel{left half} of
the figure) and \modelname{s15-B12} (\subpanel{right half}) at $t
\approx 170 ~ \ms$.  The stable downflows of the latter model
transport matter of low entropy towards the PNS where it is gains
energy by neutrino heating.  In this model, heating leads to a higher
entropy of the gas rising between the accretion columns than in model
\modelname{s15-B10}.  This is caused by a magnetic confinement of the
bubble at small radii and thus a longer exposure of the gas to strong
neutrino heating.

\begin{figure*}
  \centering
  \includegraphics[width=17cm]{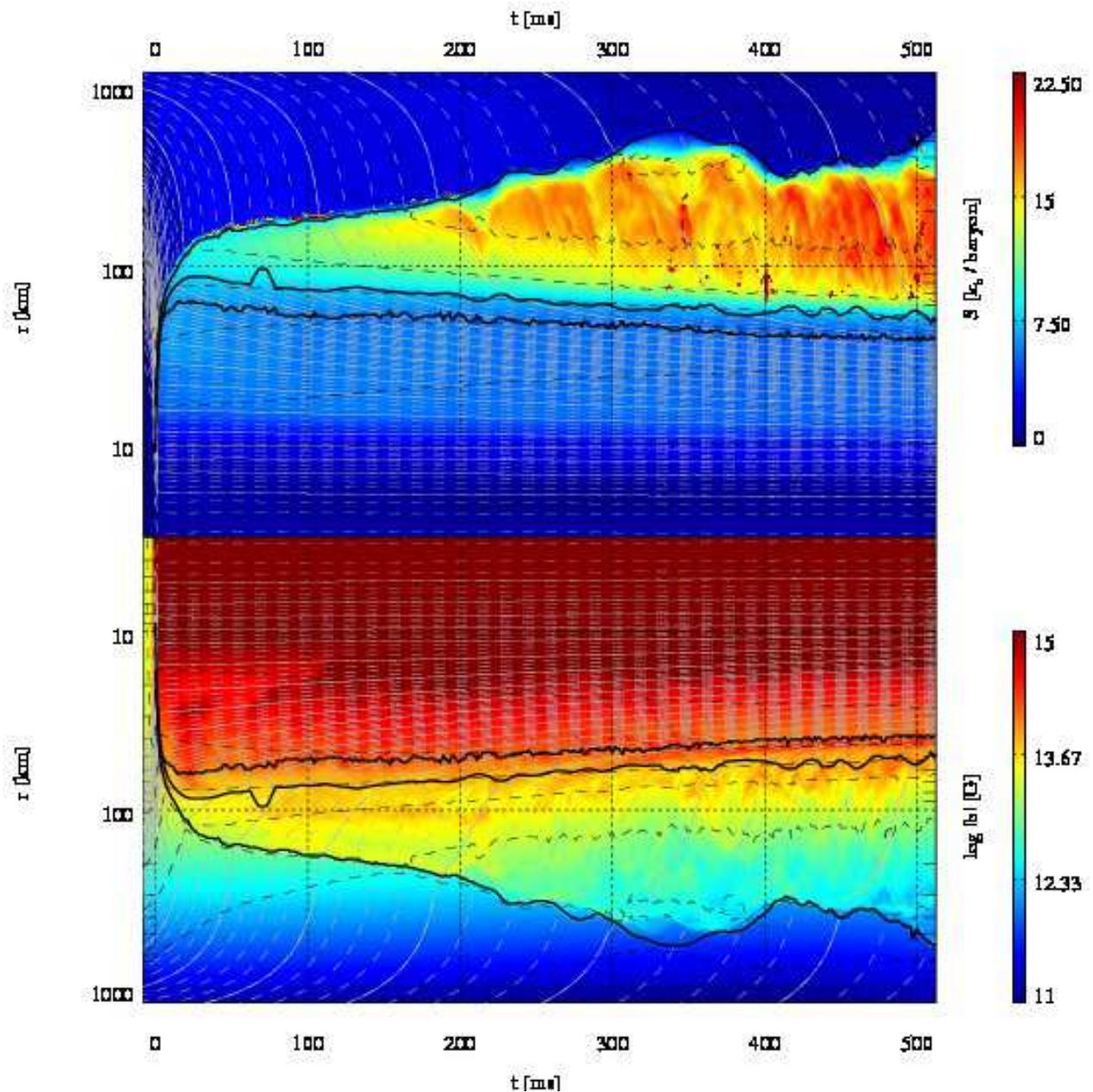}
  \caption{
    Space-time diagram of model \modelname{s15-B12}.  The top and
    bottom panels show colour-coded the entropy per baryon and the
    logarithm of the magnetic field strength, respectively.
    Additionally, mass shells and density contours are plotted, using
    the same spacing as in \figref{Fig:s15.0--Om0--B0--mshells}.  }
  \label{Fig:s15.0--Om0--B12--mshell}
\end{figure*}

The shock expands gradually until $t \approx 340 ~ \ms$ when it
reaches an outermost location of about $r \approx 600 ~ \km$.  We show
a snapshot of the model shortly before maximum shock expansion in
\figref{Fig:s15.0--Om0--B12--late2d} and a space-time diagram of the
model in \figref{Fig:s15.0--Om0--B12--mshell}.  At maximum expansion,
the quadrupolar structure is replaced by more complex flows.  The
model exhibits a large-scale pattern of low-order multipoles.  Then,
expansion ceases and the shock retreats again to $r \sim 300 ~
\mathrm{km}$, but it starts to expand again after $t \approx 420 ~
\ms$.  At the end of the simulation, $t \approx 500 ~ \ms$, its
maximum radius is 600 km again.  While large fluctuations of the shock
radius can be found in other models as well, this model is
characterised by a clearly stronger expansion.  The shock shows
appreciable deformations; however, we do not find a very large global
asphericity.  The axis ratio is around unity during the late phases of
the simulation.  This can be seen, e.g., in the space-time diagram in
\figref{Fig:s15.0--Om0--B12--mshell}: the regions of essentially high
values of the angular averaged entropy and magnetic field extend out
to the maximum shock radius.  In a model with a strongly deformed
shock, the maximum shock radius can be considerably larger than the
radius out to which these variables assume high values.

\subsection{Comparison of global quantities of our models}
\label{sSek:modcomp}

After the discussion of the magnetic-field amplification and of the
structural differences of the simulated models in the previous
sections, we conclude the presentation of our results by a comparison
of important global quantities characterising the overall evolution of
our models.

\begin{figure*}
  \centering
  \includegraphics[width=0.33\textwidth]{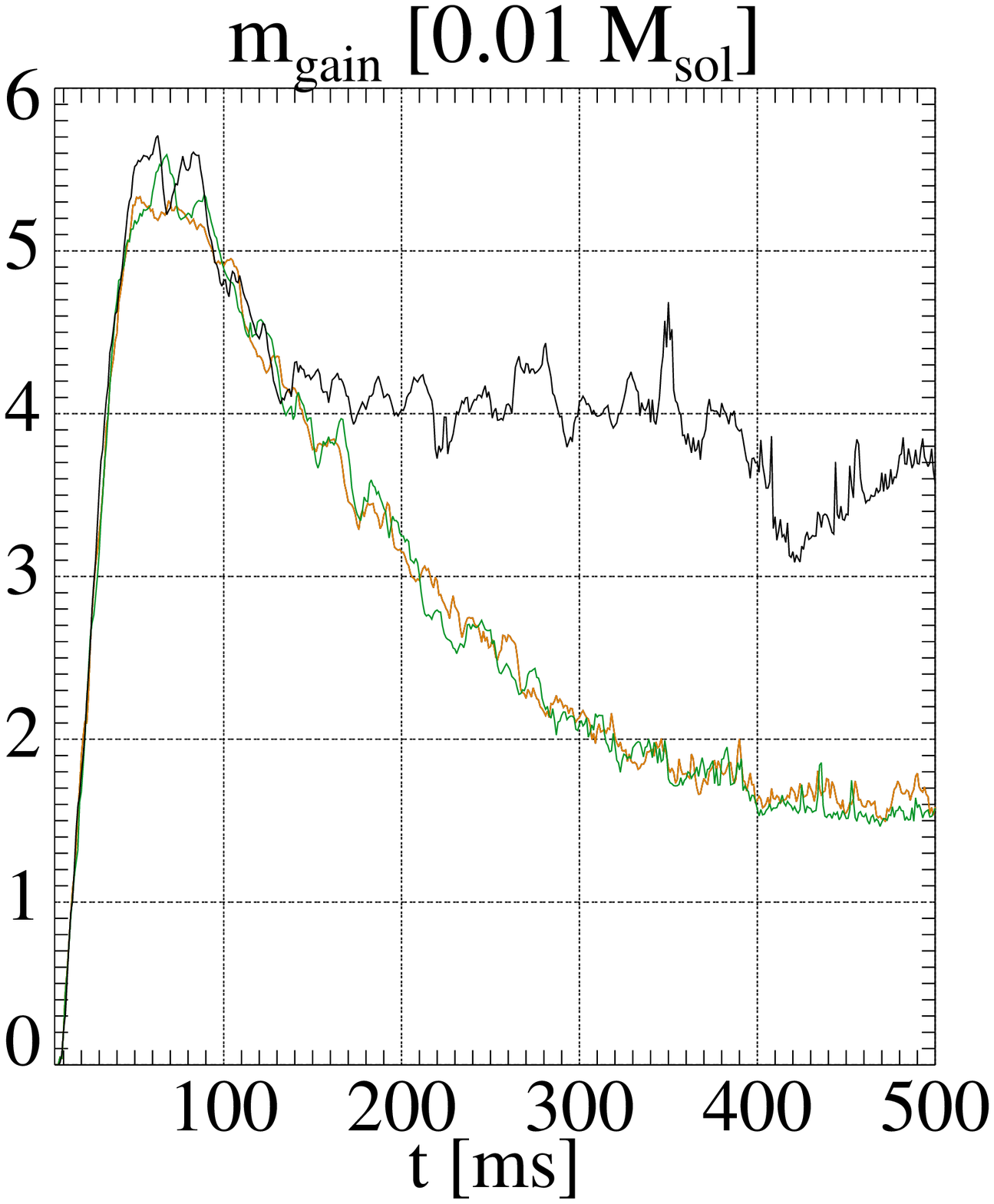}
  \includegraphics[width=0.33\textwidth]{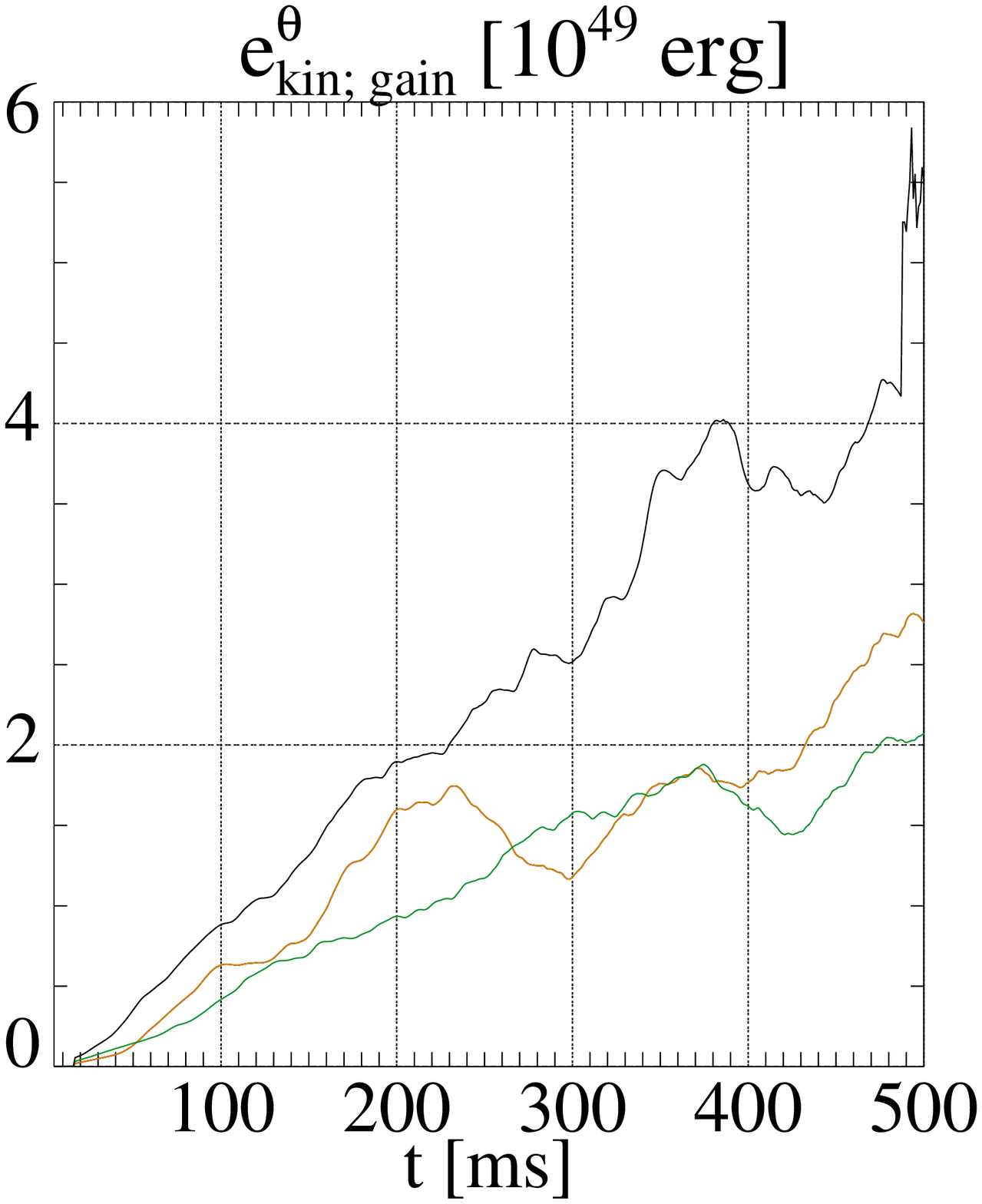}
  \includegraphics[width=0.33\textwidth]{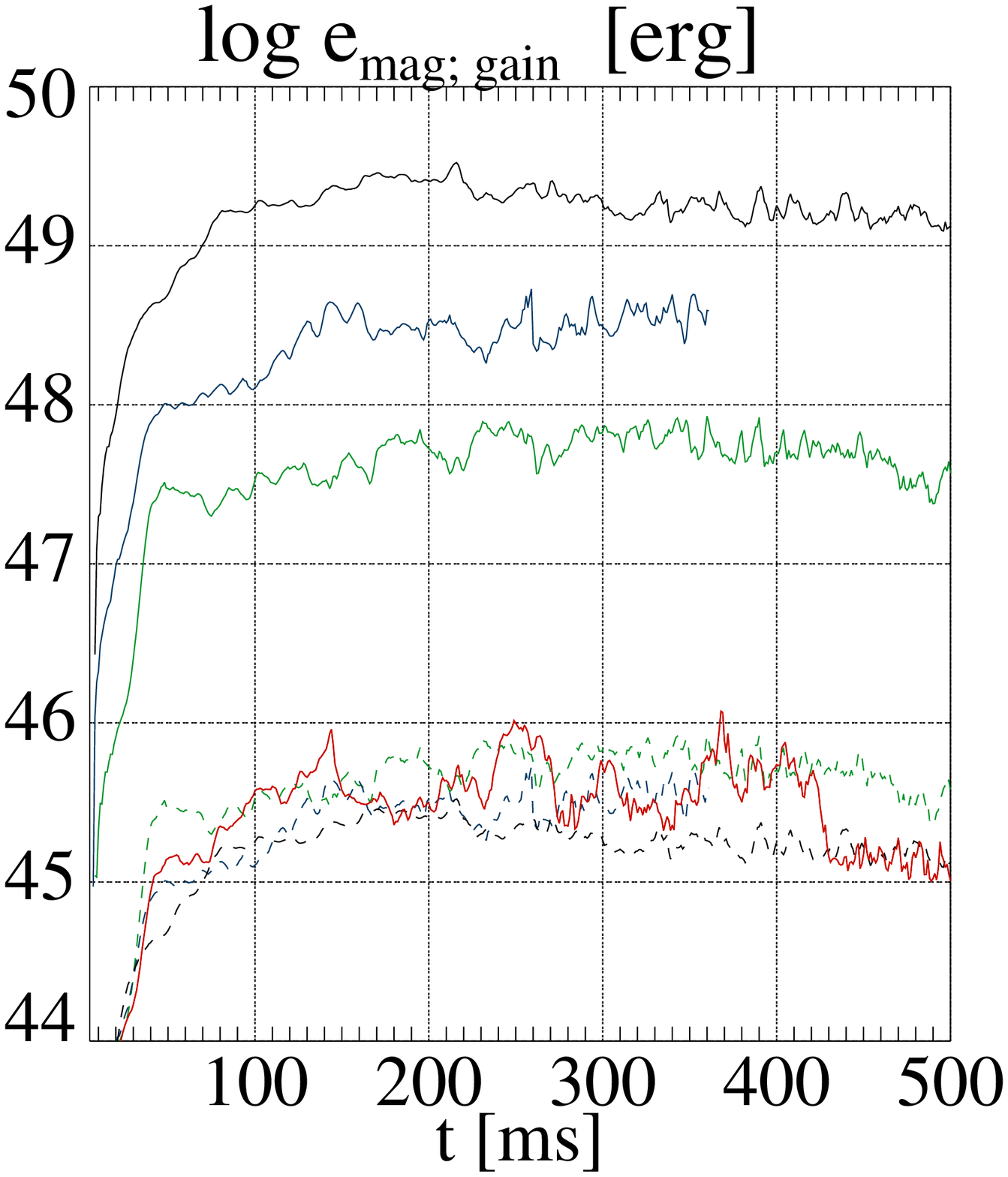}
  \caption{
    \subpanel{Left panel}: Mass in the gain region of models
    \modelname{s15-B0} (orange line), \modelname{s15-B11} (green
    line), and \modelname{s15-B12} (black line) as a function of time.
    \subpanel{Middle panel}: Lateral kinetic velocity in the gain
    region of models \modelname{s15-B0} (orange line),
    \modelname{s15-B11} (green line), and \modelname{s15-B12} (black
    line) as a function of time.
    \subpanel{Right panel}: Logarithm of the magnetic energy in the
    gain region of models \modelname{s15-B10} (red line),
    \modelname{s15-B11} (green line), \modelname{s15-B11.5} (blue
    line), and \modelname{s15-B12} (black line) as a function of time.
    The solid lines show the total magnetic energy of the
    models and the dashed lines (for the models with $b_0 > 10^{10} ~ \mathrm{G}$) the
    same quantity scaled according to the initial strength,
    respectively.  
  }
  \label{Fig:s15.0--Om0--modcomp}
\end{figure*}

We evaluate the total mass and the kinetic and magnetic energies
contained in the gain region, i.e., in the region between the shock
wave and the stable layer surrounding the PNS.  In this region, the
matter gains energy by reactions with neutrinos diffusing out of the
PNS, and hot-bubble convection and the SASI operate.

The mass in the gain region, $m_{\mathrm{gain}}$ (shown for the
non-magnetic model \modelname{s15-B0}, one weak-field model
(\modelname{s15-B11}), and the strong-field model \modelname{s15-B12}
in the \subpanel{left panel} of \figref{Fig:s15.0--Om0--modcomp}),
increases during the first 50 milliseconds post-bounce and peaks at
about $0.055 \Msol$ for all models.  Afterwards, the mass decreases
gradually, falling below $0.02 \Msol$ after $t \sim 320 ~ \ms$, in all
but the strongest magnetised model.  In the case of model
\modelname{s15-B12} $m_{\mathrm{gain}}$ instead levels off at a value
of $m_{\mathrm{gain}} \approx 0.04 \Msol$.  This different behaviour
reflects the different evolution of the shock wave.  While the
outermost shock position can reach a radius of $\sim 500 ~ \km$ along
the poles, the average shock radius oscillates around a value of
$r_{\mathrm{shock}} \gtrsim 200 ~ \km$ in the non-magnetised and the
weak-field models.  On the other hand, the shock expands to radii of
$300 - 500 ~ \km$ at all latitudes in the strong-field model.  Hence,
a higher mass can be affected by neutrino heating in the model with
the strongest initial field investigated.  Although our models do not
allow us to draw any firm conclusions on the long-term evolution of
the models, we may estimate tentatively from the evolution of
$m_{\mathrm{gain}}$ that model \modelname{s15-B12} is closest to a
successful explosion.  Further indication of a different behaviour of
weak-field and strong-field models is provided by the total energy of
the matter in the gain region, i.e., the sum of internal, kinetic,
magnetic, and gravitational energies.  We find that this total energy
of model \modelname{s15-B12} exceeds that of model \modelname{s15-B0}
by about $5 \times 10^{50} ~ \mathrm{erg}$ around $t = 440 ~
\mathrm{ms}$.

The lateral kinetic energy in the gain region, $e_{\mathrm{kin;
    gain}}^{\theta}$ (shown in the \subpanel{middle panel} of
\figref{Fig:s15.0--Om0--modcomp} for the same models), indicates the
strength of hot-bubble convection and of the SASI.  In agreement with
the evolution of $m_{\mathrm{gain}}$, the lateral kinetic energy of
model \modelname{s15-B12} exceeds that of the other models at late
times ($t \gtrsim 300 ~ \ms$) by a factor of roughly 2.  Apart from
this difference, we find a higher value of $e_{\mathrm{kin;
    gain}}^{\theta}$ in model \modelname{s15-B12} than in the other
models even at early times, $t \lesssim 120 ~ \ms$, when all models
have a very similar mass in their gain regions.  Thus, the lateral
kinetic energy per unit mass, or the average lateral velocity, is
largest in model \modelname{s15-B12}.  This is a consequence of the
peculiar quadrupolar flow pattern developing solely in this model.

We compare the amount of field amplification in the gain regions of
all magnetised models in the \subpanel{right panel} of
\figref{Fig:s15.0--Om0--modcomp}.  For all models, an initial rise of
the total magnetic energy in the gain layer $e_{\mathrm{mag; gain}}$
is followed by a phase of a relatively constant total field energies.
When taking into account the scaling of the initial fields (see the
dashed lines), we find a very similar amount of field amplification
for all models.  This result suggests that the amplification is
dominated by kinematic processes and that the back-reaction of the
field onto the flow is weak.  We do find, however, that the average
magnetic energy of model \modelname{s15-B12} is systematically lower
than expected from the scaling of the initial fields
(cf.~\figref{Fig:s15.0--Om0--modcomp}, where the black dashed line is
below the other lines).  This somewhat weaker amplification is
consistent with the presence of large subalfv\'enic regions where the
field is sufficiently strong to react back onto the flow.  The
\Alfven-wave amplification enhancing the field in this model compared
to models with weaker initial fields affects regions that are too
small to shine up in this volume-integrated energy.

Finally, we discuss the distribution of the magnetic energy in the
different regions of the core as a function of initial field strength.
We refer to \tabref{Tab:models} for a synopsis of our results.  We
compare the root-mean-square averages of the magnetic field strength
in four domains:
\begin{description}
\item[$b_{14}$] is the r.m.s.~field in the PNS at densities above
  $10^{14} ~\gccm$;
\item[$b_{\mathrm{cnv}}$] is the r.m.s.~field in the PNS convection zone;
\item[$b_{\mathrm{stb}}$] is the r.m.s.~field in the hydrodynamically
  stable region surrounding the PNS;
\item[$b_{\mathrm{gain}}$] is the r.m.s.~field in the gain (SASI)
  layer.
\end{description}
Additionally, we compare the ratios of magnetic to internal or kinetic
energy, $\beta^{\mathrm{i/k}}_{...}$, in the same regions.

For all domains, the mean fields increase monotonically with the
initial field strength, $b_0$.  The maximum field is reached in the
PNS.  Since this region is hydrodynamically stable during the
simulated evolution, amplification is dominated by radial compression.
From a weak initial field, typical pulsar fields of the order of
$10^{12-13} ~ \mathrm{G}$ can be achieved.  With a more extreme
initial field of $b_0 \sim 10^{12} ~ \mathrm{G}$ in the pre-collapse
stellar core, we get magnetar-like values of $10^{15}~ \mathrm{G}$.
Compared to the thermal energy, the magnetic field is negligible even
for this case; it can, however, dominate over the velocity field.
Since there is no feedback of the field, the final fields in the PNS
core scale very well with $b_0$.

In the PNS convection zone, the average field is somewhat weaker than
in the core of the PNS.  As discussed above, the convective velocities
can become subalfv\'enic for the strongest fields, corresponding to a
magnetic energy exceeding the kinetic energy, i.e.,
$\beta^{\mathrm{k}} > 1$; this is the case for model
\modelname{s15-B12}.  In such a situation, the efficiency of the field
amplification is limited.  Consequently, the scaling of the
r.m.s.~field with $b_0$ breaks down for $b_0 \gtrsim 3 \times 10^{11}
~ \mathrm{G}$.  The field found for model \modelname{s15-B12},
$b_{\mathrm{cnv}} \approx 4.7 \times 10^{14} ~ \mathrm{G}$, should be
close to the maximum field that can be achieved by the PNS convection
excited in this core.

Similar results can be found in the hydrodynamically stable layer
between the PNS convection zone and the gain region.  The
r.m.s.~magnetic field strength is weaker by a factor of $\sim 3$.
Again, we find a ratio of magnetic to kinetic energy in excess of
unity for the strongest initial field considered and a deviation of
the final magnetic field strength from a simple scaling with the
initial field strength.

The r.m.s.~field strength in the gain layer is between $3.6 \times
10^{11} ~ \mathrm{G}$ for model \modelname{s15-B10} and $1.2 \times
10^{13} ~ \mathrm{G}$ for model \modelname{s15-B12}.  In all models,
$\beta^{\mathrm{k}}$ is below unity, though.  As discussed in
\secref{ssSek:Acccol} and \secref{sSek:Res-Bs}, trans- and
subalfv\'enic regions are present and have an important influence on
the dynamics and the field amplification.  Because the magnetic field
strength decreases more slowly with radius than the internal energy,
$\beta^{\mathrm{i}}$ can be as large as one per cent here.

Finally, we note that for all models by far most (of the order of two
thirds) of the magnetic energy is contained in the PNS convection
region.  The magnetic energies in the stable layer are of the same
order of magnitude as those in the gain region, though the exact value
of $b_{\mathrm{gain}}$ is variable and depends strongly on the size of
the gain region.


\section{Summary and conclusions}
\label{Sek:SumCon}

We have studied (some of) the processes leading to the amplification
of the magnetic field in a non-rotating stellar core during the
collapse and post-bounce accretion phases of a supernova.  In
non-rotating stars, a variety of amplification mechanisms considered
in previous works, e.g., winding of the field by differential rotation
or the MRI, are not viable.  Instead, convection and the SASI may
constitute (small-scale) dynamos.  Additionally, \Alfven~waves
travelling against the gas flow may be amplified once they reach an
\Alfven~point where the (co-moving) \Alfven~velocity equals the gas
velocity, a condition that could be fulfilled in the accretion flow
onto the PNS.  If efficient amplification occurs, the field may be
able to affect the dynamics of the core, e.g., by altering the
geometry of SASI flows or by energy dissipation through \Alfven~waves
in the upper layers of the hot-bubble region.

Reducing the complexity of the problem, simplified models, e.g.,
one-dimensional simulations with assumptions for the excitation,
propagation and dissipation of \Alfven~waves
\citep{Suzuki_Sumiyoshi_Yamada__2008__ApJ__Alfven_driven_SN}, toy
models of \Alfven~waves in decelerating flows
\citep{Guilet_et_al__2010__ArXive-prints__Dynamics_of_an_Alfven_surface_in_CCSNe},
and 2D and 3D MHD simulations without neutrino transport
\citep[][]{Endeve_et_al__2010__apj__Generation_of_Magnetic_Fields_By_the_SASI}
have demonstrated that these effects could, in principle, be relevant.
On the other hand, to study their evolution under less idealised
conditions, in particular their interplay with a highly dynamical
background, more calculations of self-consistent models are required,
i.e.~multi-dimensional MHD including a treatment of neutrino transfer
through the stellar core.

To this end, we have performed axisymmetric simulations of the
collapse and the post-bounce evolution of the core of a star of 15
solar masses possessing a purely poloidal initial field.  Using a new
code for neutrino-magnetohydrodynamics, we have solved the MHD
equations coupled to the system of two-moment equations for the
neutrino transport; the closure for the moment equations was provided
by an analytic variable Eddington factor.  We included descriptions
for the most important reactions between electron neutrinos and
antineutrinos and the stellar matter, viz.~nucleonic and nuclear
emission and absorption and scattering off nucleons and nuclei.
Compared to current state-of-the-art techniques such as Boltzmann
codes, our approach ignores muon and tau neutrinos and is less
accurate in treating neutrino-matter interactions, but needs much less
computational resources and possesses full two-dimensionality
including velocity effects.  We are able to reproduce the basic
features found in detailed simulations, e.g., the stagnation of the
prompt shock wave, PNS and hot-bubble convection, and the SASI
activity.  Hence, our models allow for a fairly reliable assessment of
the main MHD effects in self-consistent SN core models; we discuss the
remaining major limitations below.

The principal results of our simulations and the main conclusions to
be drawn from our results are:
\begin{enumerate}
\item The magnetic field is amplified kinematically by the turbulent
  flows developing due to convection and the SASI.  The amplification
  factor does not depend on the initial field unless it reaches,
  starting from a high initial value, equipartition with the kinetic
  energy, the maximum possible energy attainable.  The maximum field
  we observe is a few times $10^{15} ~ \mathrm{G}$ at the base of the
  PNS convection layer.  In the PNS convection zone, the radial
  profile of the field strength can be approximated by a power law,
  $|b| \propto r^{-2}$.
\item When falling through the non-spherical supernova shock, the
  magnetic field is bent by lateral flows, creating a component
  parallel to the shock.  These perturbations are advected towards the
  PNS convection zone.  The field accumulates there, leading to a
  layer of strong magnetic fields.  Since the perturbations are
  associated with a strong lateral component of the field, this layer
  is dominated by the $\theta$-component of the field.  Hence,
  \Alfven~waves propagate at constant radius rather than upwards.
  Therefore, we do not find an \Alfven~surface in the accretion flow,
  although there are sub- as well as superalfv\'enic regions in this
  layer.  This limits the efficiency of the amplification mechanism
  proposed by
  \cite{Guilet_et_al__2010__ArXive-prints__Dynamics_of_an_Alfven_surface_in_CCSNe}.
\item Conditions for the latter effect are more favourable if
  accretion occurs through a rather stationary column.  In our
  axisymmetric models, this is the case mostly along the polar axis
  where the geometry enforces radial fields and flows.  An
  \Alfven~surface forms in the radial field of the accretion column at
  a radius depending on the initial field strength.  For fields of
  $b_0 \gtrsim 10^{11} ~ \mathrm{G}$ initially, we observe that
  perturbations created in the PNS convection zone propagate along the
  field upwards into the accretion column.  Interacting with the
  accretion flow, they lead to a steepening of the profile of the
  accretion velocity near the lower end of the accretion colum to a
  discontinuity where a strong enhancement of the magnetic field takes
  place.
\item The strongest initial field we have investigated, $b_0 = 10^{12}
  ~ \mathrm{G}$, is able to shape the post-shock flow.  It favours the
  formation of a flow dominated by low-order multipole modes, viz.~a
  quadrupolar pattern of accretion columns at the poles and near the
  equator.  In the stable polar columns we observe, as in the case of
  lower fields ($b_0 = 10^{11} ~ \mathrm{G}$), an interaction of
  perturbations travelling upwards from the PNS convection zone with
  the accretion flow.  This gives rise to the formation of a layer of
  very strong field.  Also in the convective and SASI layers behind
  the shock, the field strength grows with the magnitude of the
  initial field.  In this model with its particular flow topology, we
  find the most pronounced shock expansion of all models.
\end{enumerate}

In summary, our results suggest that magnetic field amplification to
interesting strengths can efficiently take place during the stellar
core collapse even in the absence of rotation.  In addition to an
enhancement due to compression by the radial collapse, we find that
non-radial fluid flows associated with convection and SASI activity,
and the interaction of \Alfven~waves in the accretion flows can
amplify the initial iron-core fields.  Starting with
$10^{9}$--$10^{10}$\,G as predicted by present stellar evolution
models
\citep[][]{Heger_et_al__2005__apj__Presupernova_Evolution_of_Differentially_Rotating_Massive_Stars_Including_Magnetic_Fields},
field strengths of typical pulsars ($10^{12}$--$10^{13}$\,G) can be
reached. Magnetar fields of $10^{14}$--$10^{15}$\,G result when the
progenitor core possesses a pre-collapse field between a few
$10^{11}$\,G and $10^{12}$\,G. Only in the latter case, the fields
around the nascent neutron star obtain dynamical importance and might
have an influence on the supernova explosion mechanism.

Though offering some insight into the magnetic-field evolution in
non-rotating magnetised cores, our study has several important
limitations:
\begin{enumerate}
\item We have used a new treatment of neutrino transport and
  simplified neutrino-matter interactions constraining ourselves to
  electron neutrinos and antineutrinos.  While the accuracy of the
  transport compared to other schemes has to be assessed in a separate
  study, and our set of neutrino reactions is certainly insufficient
  for high-precision supernova physics, we do not deem this a problem
  for the presented investigation because we are able to capture the
  most important dynamical effects in a supernova core.
\item To save computational costs, we have restricted ourselves to
  axisymmetric simulations.  In the light of the anti-dynamo theorems,
  this is a severe limitation leading to a wrong, possibly too low,
  level of amplification of the field in the turbulent regions.
  Furthermore, our models do not allow for the development of shear
  layers associated with non-axisymmetric spiral modes of the SASI,
  which may also be a site of efficient amplification of the magnetic
  field \citep[see
  also][]{Endeve_et_al__2010__apj__Generation_of_Magnetic_Fields_By_the_SASI}.
\item Moreover, axisymmetry restricts the dynamics of the accretion
  flows, favouring the development of very stable accretion columns
  along the poles.  As we have seen, field amplification shows very
  distinct features in and below these columns.  We presume that the
  dynamics in three dimensions is dominated by what we have seen in
  off-axis accretion flows, i.e., less stable \Alfven~surfaces and
  less efficient amplification of \Alfven~waves.  Very strong initial
  fields of the order of $b_0 = 10^{12} ~ \mathrm{G}$ lead to high
  field strengths that are able to dominate the post-shock accretion
  flow.  This may establish coherent, stable accretion columns.  For
  such fields, our axisymmetric results may hence be a decent
  approximation.
\item Turbulent field amplification may be very sensitive to
  dissipation coefficients, physical and numerical.  Our models, based
  on \subpanel{ideal} MHD, neglect dissipation by physical viscosity
  and resistivity, but are computed on relatively coarse numerical
  grids corresponding to excessive numerical dissipation.  Therefore,
  we are not able to follow the turbulent (inverse) cascades of
  magnetic and kinetic energy and helicity covering many orders of
  magnitude in wave number in a supernova core.  The effect of
  insufficient resolution on \Alfven~waves is probably less serious
  though their wave number should increase as they approach the an
  \Alfven~point, requiring enhanced resolution.  Simulations of cores
  at a resolution corresponding to numerical viscosity and resistivity
  below the physical ones are by far too expensive today, and will
  remain so for a long time.  To tackle this difficulty, a combination
  of different approaches would be desirable, viz.~global direct
  numerical simulations of the core with drastically enhanced physical
  transport coefficients, and simple sub-grid models for MHD
  turbulence based on idealised local simulations neglecting most
  aspects of, e.g., neutrino physics.  We are, however, aware of the
  lack of reliable sub-grid models for MHD at present, obstructing
  further progress in this direction.
\end{enumerate}

Apart from these methodological shortcomings, open physical questions
are, e.g., the influence of the progenitor on the establishment of
certain patterns in the accretion flow, effects of very strong initial
fields, and the influence of slow rotation of the core on our
findings.  In particular the last issue may prove interesting as it
would enable a large-scale dynamo.  We defer these questions as well
as the more technical problems listed above to further investigations.

\begin{acknowledgements}
  HTJ is grateful to J{\'e}r{\^o}me Guilet and Thierry Foglizzo for
  interesting and educative discussions.  M.O.~is supported by a
  \emph{Golda Meir Postdoctoral Fellowship} of the Hebrew University
  of Jerusalem.  This work was supported by the Deutsche
  Forschungsgemeinschaft through the Transregional Collaborative
  Research Centers SFB/TR~27 ``Neutrinos and Beyond'' and SFB/TR~7
  ``Gravitational Wave Astronomy'' and the Cluster of Excellence
  EXC~153 ``Origin and Structure of the Universe''
  (\texttt{http://www.universe-cluster.de}). The simulations were
  performed at the Rechenzentrum Garching (RZG) of the Max Planck
  Society.
\end{acknowledgements}

\appendix

\section{Neutrino transport}
\label{Sek:App-NT}

Most results on magnetic effects in supernovae draw from more or less
idealised investigations, e.g., analytic calculations or simplified
models focusing on one particular aspect of the problem while
neglecting some of the complex physics of the explosion and, thus,
limiting their range of application.  Due to the enormous
computational requirements of an accurate treatment of the transport
of neutrinos and their interaction with matter, simplifications are
most commonly made in this sector.  Not discussing local simulations
focusing on a high-resolution modelling of a small part of the flows,
we can distinguish between different approximations:
\begin{enumerate}
\item Some studies
  \citep[e.g.,][]{Obergaulinger_Aloy_Mueller__2006__AA__MR_collapse,Mikami_et_al__2008__apj__3d_MHD_SN}
  neglect the effect of neutrinos completely and use only a
  simplified, e.g., polytropic, equation of state.  Typically, these
  models exhibit prompt explosions, limiting their validity to the
  immediate (few milliseconds) post-bounce phase.  Even without any
  treatment of neutrinos, it is possible to simulate the development
  of the SASI
  \citep[e.g.,][]{Blondin_Mezzacappa_DeMarino__2003__apj__Stability_of_Standing_Accretion_Shocks_with_an_Eye_toward_Core-Collapse_Supernovae}.
\item A way to model the dynamics qualitatively correctly while
  avoiding the complications of a full neutrino-transport scheme is
  the use of simple heating/cooling or deleptonisation source terms.
  Such schemes prescribe the loss and gain of energy or electron
  number by cooling and heating terms that depend on the local
  thermodynamic state of the matter (e.g., density, temperature,
  electron fraction).  Depending on the details of the model, the can
  reproduce, e.g., the correct deleptonisation behaviour during
  collapse
  \citep[e.g.,][]{Liebendorfer__2005__apj__Deleptonisation_scheme} or
  the development of the SASI
  \citep[][]{Fernandez_Thompson__2009__apj__Stability_of_a_Spherical_Accretion_Shock_with_Nuclear_Dissociation,Fernandez__2010__ArXive-prints__The_Spiral_Modes_of_the_SASI}.
\item A leakage scheme accounts for the optical depth of the region
  emitting neutrinos is taken into account in the cooling terms.  The
  results of such simulations can be similar to that of models with
  purely local cooling terms
  \citep[e.g.,][]{Kotake_etal__2004__Apj__SN-magrot-neutrino-emission}.
\item Despite the large computational effort, various methods for
  energy-dependent neutrino hydrodynamics have been used;
  state-of-the-art codes include, e.g., the \emph{isotropic diffusion
    source approximation}, a scheme decomposing the neutrinos into a
  trapped, diffusing, and a freely streaming component
  \citep[][]{Liebendorfer_et_al__2009__apj__The_Isotropic_Diffusion_Source_Approximation_for_Supernova_Neutrino_Transport},
  flux-limited diffusion (FLD), or solve the Boltzmann equation for
  the neutrino phase-space distribution.  Among the codes based on the
  FLD method, there is a wide variety of approaches to the treatment
  of multi-dimensional effects and the coupling of energy bins:
  \begin{itemize}
  \item 
    \cite{Livne_et_al__2004__apj__VULCAN_2d_Multigroup_Multiangle_RHD_Test_Simulation_in_the_CCSN_Context,Burrows_etal__2007__ApJ__MHD-SN}
    use a multi-group FLD code without energy-bin coupling and no
    velocity terms in the moments equations,
  \item
    \cite{Walder_et_al__2005__apj__Anisotropies_in_the_Neutrino_Fluxes_and_Heating_Profiles_in_2d_MGRHD_rotating_CCSNe,Ott_et_al__2008__apj__2d_multi-angle_MGFLD_SN}
    use a two-dimensional multi-angle, multi-group FLD code without
    energy-bin coupling and no velocity terms in the moment
    equations,
  \item the simulations of
    \cite{Mezzacappa_et_al___2007__TheMulticoloredLandscapeofCompactObjectsandTheirExplosiveOrigins__Ascertaining_the_Core_Collapse_Supernova_Mechanism:_An_Emerging_Picture?,Yakunin_et_al__2010__CQGra__GW_from_CCSNe}
    employ a \emph{ray-by-ray-plus} multi-group FLD method that
    includes the energy-bin coupling terms and the velocity terms up
    to order $\mathcal{O}(v/c)$.
  \end{itemize}
  Similar distinctions can be made for the codes based on Boltzmann solvers:
  \begin{itemize}
  \item
    \cite{Mezzacappa_Bruenn__1993__apj__Solving_neutrino_Boltzmann_equation_coupled_to_spherical_core_collapse,Mezzacappa__2001__PRL__Spherical_Core_Collapse_Bounce_and_Postbounce_Evolution_Boltzmann_Neutrino_Transport,Liebendorfer_et_al__2004__apjs__FD_Neutrino_GRRHD_Agile-Boltztran}
    solve the one-dimensional Boltzmann equation by an $S_n$
    technique,
  \item \cite{Rampp_Janka__2002__AA__Vertex} employ a
    one-dimensional two-moment-closure scheme with a closure
    relation obtained from the solution of a model Boltzmann equation,
  \item this scheme was generalised to a ray-by-ray-plus version by
    \cite{Buras_etal__2006__AA__Mudbath_1}, accounting for all terms
    coupling energy bins and all velocity-dependent terms.
    \cite{Mueller_et_al__2010__apjs__A_New_Multi-dimensional_General_Relativistic_Neutrino_Hydrodynamic_Code_for_Core-collapse_Supernovae.I.Method_and_Code_Tests_in_Spherical_Symmetry}
    developed a 2D general relativistic hydro solver with a
    ray-by-ray-plus neutrino transport similar to
    \cite{Buras_etal__2006__AA__Mudbath_1}.
  \end{itemize}
  These schemes provide the best models of the complex
  flow patterns of a supernova and allow for a reliable distinction
  between successful and failed supernovae and an accurate prediction
  of the explosion energy.  Testing the limits of currently available
  supercomputers, two-dimensional axisymmetric simulations are
  possible, but the step to three-dimensional models may require a
  profound change of algorithms and programming models to take full
  advantage of the next generation of computers.
\end{enumerate}

Neutrino transfer describes the neutrinos in terms of their
phase-space distribution function, $f (t, \vec x, \epsilon, \vec n)$,
a function of time, $t$, position, $\vec x$, neutrino energy,
$\epsilon$, and propagation direction, $\vec n$ ($|\vec n| = 1$).  The
dimensionality of this function is the main reason for the high
computational requirements of RT.  It can be reduced by evolving the
set of neutrino moments, $\neutrino{M}^k = \int \mathrm{d} \Omega \,
\vec n^kf$, $k=0,...,\infty$, rather than the distribution function;
the first moments have an immediate physical meaning: $k=0,1,2$
correspond to the neutrino energy density, momentum density, and
pressure tensor, respectively.

In principle, the full infinite series of moments is required for an
accurate representation of $f$.  In practise, the series is truncated
at some level, $k_{\mathrm{max}}$, retaining the conservation laws for
the first few moments.  Since the equation for the $k^{\mathrm{th}}$
moment,
\begin{equation}
  \label{Gl:RT-moments}
  \partial_{t} \neutrino{M}^k + \vec \nabla \neutrino{M}^{k+1}
  = \neutrino{Q}^k,
\end{equation}
involves at least the next moment (order $k+1$), a closure is required
to relate the first moment beyond truncation,
$\neutrino{M}^{k_{\mathrm{max}}+1}$, to the moments of order $0, ...,
k_{\mathrm{max}}$ (the source term, $\neutrino{Q}^k$, accounts for all
interaction terms and for velocity-dependent terms such as advection,
compression, or Doppler shifts).  A common choice is the
$0^{\mathrm{th}}$-moment system, the equation for the neutrino energy,
with the first moment given by the (flux-limited) diffusion flux.  We
go one step beyond this and evolve the $1^{\mathrm{st}}$-moment
system,
\begin{eqnarray}
  \label{Gl:RT-0mom-1}
  \partial_{t} \neutrino{E} + \vec \nabla \neutrino{F} 
  & = & \neutrino{Q}^0,
  \\
  \label{Gl:RT-1mom-1}
  \partial_{t} \neutrino{F} + \vec \nabla \neutrino{P} 
  & = &  \neutrino{\vec Q}^1,
\end{eqnarray}
closing the system by assuming a simple analytic form of the
$2^{\mathrm{nd}}$ moment, $\neutrino{P} = \neutrino{P}
(\neutrino{E},\neutrino{F})$.  In a one-dimensional system, one scalar
function is sufficient to define the closure, the \emph{Eddington
  factor} $\neutrino{p}$.  Following
\cite{Audit_et_al__2002__astro-ph__hyp_RHD_closure}, we construct a
tensorial closure for multi-dimensional systems from the
one-dimensional Eddington factor:
\begin{equation}
  \label{Gl:RT-multiD-Eddington}
  \neutrino{P}^{ij} 
  = 
  \left(
    \frac{1 - \neutrino{p}}{2} \delta^{ij} 
    + \frac{3 \neutrino{p} - 1}{2} \frac{\neutrino{F}^i
      \neutrino{F}^j}{\neutrino{F}^2}
    \right)
    \neutrino{E}
    .
\end{equation}
This expression combines a diagonal part, accounting for isotropic
diffusion, with a contribution accounting for free streaming in the
direction of the neutrino flux.  In the diffusion limit, the Eddington
factor is $\neutrino{p} = 1/3$, and the pressure tensor is given by
the isotropic part alone, $\neutrino{P}^{ij} = \neutrino{E}/3
\delta^{ij} $, and in the free-streaming limit, $\neutrino{p} = 1$,
only the second part contributes, the total neutrino pressure is
$\neutrino{E}$ and $\neutrino{P}^{ij}$ is peaked in the propagation
direction of the neutrinos.
\cite{Pons_Ibanez_Miralles__2000__MNRAS__hyperbol_radtrans} have
demonstrated the applicability of standard HRSC methods to this
(hyperbolic) system and compared different suggestions for the
closure.  In our simulations, we use the Eddington factor by
\cite{Minerbo__1978__jqsrt__Maximum_entropy_Eddington_factors} To
avoid the evaluation of the Langevin function in the expression for
the closure, we apply a simple polynomial approximation given by
\cite{Cernohorsky_Bludman__1994__ApJ__MEC-Transport}:
\begin{equation}
  \label{Gl:RT-Minerbo}
  \neutrino{p} 
  = \frac{1}{3} 
  + \frac{1}{15} 
  \left( 6 \neutrino{f}^2 - 2 \neutrino{f}^3 + 6 \neutrino{f}^4
  \right),
\end{equation}
where $\neutrino{f} = \neutrino{F} / c \neutrino{E}$ is the flux
factor of the neutrinos, i.e., the neutrino flux normalised to the
maximum flux allowed by causality.

Exploiting the hyperbolic character of the RT system, we use HRSC
methods for the moments equations, viz.~the same high-order
reconstruction as in the MHD subsystem, either the Lax-Friedrichs or
the HLL approximate Riemann solver, and an explicit Runge-Kutta time
integrator.  This approach meets its limitations when the Peclet
number, $\mathrm{Pe} = \chi^1 \delta x$, i.e., the optical depth of
one grid cell of width $\delta x$ ($c$ is the speed of light), exceeds
one.  Then, the interaction source term of the neutrino fluxes, $- c
\chi^1 \neutrino{F}$, becomes stiff (the characteristic time scale
being $\sim 1 / (c \chi)$), enforcing a time step much below that of
the hyperbolic MHD or transport parts, $(\delta t)_{\mathrm{hyp}} \leq
\delta x / c_{\mathrm{max}}$, for a given maximum characteristic speed
$c_{\mathrm{max}}$.  Similarly to
\cite{Pons_Ibanez_Miralles__2000__MNRAS__hyperbol_radtrans} and
\cite{Audit_et_al__2002__astro-ph__hyp_RHD_closure}, we modify the
Riemann fluxes of the $0^{\mathrm{th}}$ moments to reproduce the
correct parabolic diffusion limit and use an implicit time integrator
for the interaction source term of the first moment, thereby avoiding
a time-step restriction below $(\delta t)_{\mathrm{hyp}}$.  A stiff
part of the system on its own if $c \chi^0 \delta t > 1$, the
neutrino-matter interaction source terms in the
$0^{\mathrm{th}}$-order moment equation are treated implicitly as
well.

Including contributions due to the fluid velocity of order
$\mathcal{O} (v/c)$, we solve the equations for the neutrino moments
in the \emph{co-moving} frame of reference, summarised, e.g., by
\citet{Munier_Weaver__1985__CoPhr__Comoving_Radtrans_2} ($\omega$ is
the neutrino energy):
\begin{eqnarray}
  \label{Gl:RT-cmv0}
  \partial_{t} \neutrino{E} 
  + \vec \nabla \cdot ( \neutrino{E} \vec v + \neutrino{\vec F})
  - \omega \nabla_j v_k \partial_{\omega} \neutrino{P}^{jk}
  & = &
  \neutrino{S}^0
  ,
  \\
  \label{Gl:RT-cmv1}
  \partial_{t} \neutrino{F}^i 
  + \nabla_j ( \neutrino{F}^i v^j + \neutrino{P}^{ij} )
  + \neutrino{F}^i \nabla_j v^j
  & = &
  \neutrino{S}^{1; i}
  .
\end{eqnarray}
In these equations, we have separated the total source terms,
$\neutrino{Q}^{0,1}$ (\eqref{Gl:RT-0mom-1} and \eqref{Gl:RT-1mom-1}),
into source terms due to neutrino-matter reactions,
$\neutrino{S}^{0,1}$, and velocity-dependent terms.  We ensure the
simultaneous conservation of neutrino energy and number density by
evolving a duplicate set of neutrino moments for both conserved
variables.  There are alternatives to this approach, e.g., using a
consistent discretisation of the neutrinos in energy space
\citep[][]{Mueller_et_al__2010__apjs__A_New_Multi-dimensional_General_Relativistic_Neutrino_Hydrodynamic_Code_for_Core-collapse_Supernovae.I.Method_and_Code_Tests_in_Spherical_Symmetry}.

Striving for simplification of our modelling, we include a reduced set
of neutrino-matter interactions only.  It consists of 
\begin{enumerate}
\item emission and absorption of electron neutrinos by neutrons,
\item emission and absorption of electron anti-neutrinos by protons,
\item elastic scattering of all neutrino flavours off nucleons,
\item emission and absorption of electron neutrinos by heavy nuclei,
\item coherent elastic scattering of all neutrino flavours off heavy nuclei.
\end{enumerate}
The reaction rates and opacities are implemented following
\cite{Rampp_Janka__2002__AA__Vertex}.  Neglecting pair processes, we
do not create $\mu$ and $\tau$ neutrinos.  Furthermore, we neglect
inelastic scattering of neutrinos off leptons.  From these reactions,
we can construct the source terms of the $0^{\mathrm{th}}$ and
$1^{\mathrm{st}}$ moments,
\begin{eqnarray}
  \label{Gl:RT-src0}
  \neutrino{S}^0 &= & c \chi^0 ( \neutrino{E}^{\mathrm{eq}} -
  \neutrino{E}),
  \\
  \label{Gl:RT-src1}
  \neutrino{S}^1 &= & - c \chi^1 \neutrino{F},
\end{eqnarray}
where $\neutrino{E}^{\mathrm{eq}}$ is the equilibrium (Fermi-Dirac)
value of the $0^{\mathrm{th}}$ moment.  The opacities for the
$0^{\mathrm{th}}$ and $1^{\mathrm{st}}$ moments include absorption and
emission processes only and absorption, emission, and scattering
processes, respectively.

The number of neutrinos in a certain energy bin is limited by the
Pauli exclusion principle.  If a bin is populated by a number of
neutrinos close to this limit, interactions, in particular inelastic
scattering, redistributing the particles in energy space, are getting
phase-space blocked such that a violation of the Pauli principle is
avoided.  Numerical schemes solving the interaction terms and the
spatial derivatives of the transport equations time-implicitly, obey
this constraint naturally.  Since we are using an explicit time
integrator for the spatial derivatives instead, we have to enforce
this constraint manually similarly to
\cite{Swesty_Myra__2005__OpenIssuesinCoreCollapseSupernovaTheory__Advances_in_Multi-Dimensional_Simulation_of_Core-Collapse_Supernovae}.

\bibliographystyle{aa}
\bibliography{../biblio.bib}

\end{document}